\documentclass[aps,prd,preprint,showpacs,nofootinbib,floatfix]{revtex4-1}
\usepackage{amssymb}
\usepackage{amsmath}
\usepackage{amsfonts}
\usepackage{graphicx}
\usepackage{natbib}

\graphicspath{{figures/}}

\begin{document}


\title{Red Spectral Tilt and Observable Gravity Waves in Shifted Hybrid Inflation}

\author{Matthew Civiletti}\email{mcivil@udel.edu}
\author{Mansoor Ur Rehman}\email{rehman@udel.edu}
\author{Qaisar Shafi}\email{shafi@bartol.udel.edu}
\author{Joshua R. Wickman}\email{jwickman@udel.edu}
\affiliation{Bartol Research Institute, Department of Physics and Astronomy, 
University of Delaware, Newark, Delaware 19716, USA}

\begin{abstract}
We consider supersymmetric shifted hybrid inflation models with a red tilted scalar spectral index $n_s$ in agreement with the WMAP 7-yr central value.  If non-minimal supergravity corrections are included, these models can also support a tensor-to-scalar ratio as large as $r \simeq 0.02$, which may be observable by the Planck Satellite. In contrast to the standard supersymmetric hybrid inflation scenario, topological defects produced via gauge symmetry breaking are inflated away in the shifted version of the theory.
\end{abstract}


\maketitle


\section{Introduction}

With the Large Hadron Collider (LHC) beginning to yield data, these are exciting times in particle physics.  It is anticipated that the LHC will definitively answer the question of whether supersymmetry (SUSY), the most highly sought and motivated exension of the Standard Model to date, is present at energies as low as the TeV range.  As if this were not enough, cosmologists are on the edge of our proverbial seats as we await the results of the Planck Satellite mission, launched nearly two years ago and scheduled to release new precision results early next year.  Both mainstream particle physics and inflationary cosmology are simultaneously poised to undergo great strides in the near future, and inflationary models that make use of LHC-testable ideas such as SUSY are at the bleeding edge.

One such class of models, lying at this interface between cosmology and mainstream particle physics, is SUSY hybrid inflation~\cite{Dvali:1994ms, Copeland:1994vg}.  Current events aside, models of SUSY hybrid inflation are well-motivated in a variety of other ways.  This framework naturally incorporates grand unified theories (GUTs)~\cite{Senoguz:2003zw}; the SUSY hybrid potential for the scalar components of the superfields constitutes a supersymmetric extension of the Higgs potential, through which the GUT symmetry breaking is achieved.  SUSY is temporarily broken during inflation and subsequently restored at the vacuum expectation value (vev) of the system, allowing for the usual mechanisms of SUSY breaking to be employed at low energies.  A natural extension of these models is to generalize SUSY to its local form, supergravity (SUGRA)~\cite{Panagiotakopoulos:1997qd, *Linde:1997sj, *Buchmuller:2000zm, *Kawasaki:2003zv}.

In a series of recent calculations, we have shown that the `standard' edition of SUSY hybrid inflation models can be brought into good agreement with the latest experimental data from the WMAP 7-yr analysis~\cite{Komatsu:2010fb} if the model is generalized to include various well-motivated corrections to the supergravity potential~\cite{Senoguz:2004vu, Jeannerot:2005mc}.  If suitable soft SUSY-breaking terms are carried over from the hidden sector, only the canonical form of the K\"ahler potential is needed to produce a red-tilted spectrum with a scalar spectral index $n_s$ that can lie anywhere in the 2$\sigma$ region~\cite{Rehman:2009nq, Rehman:2009yj, *[{For a reduced spectral index using non-minimal K\"ahler, see }]  [{.}] Pallis:2007du}.  If the model is extended to include non-minimal terms in the K\"ahler potential~\cite{BasteroGil:2006cm, urRehman:2006hu}, there is also a region of parameter space where large tensor modes are possible, with the tensor-to-scalar ratio $r$ reaching as high as 0.03, potentially measurable by Planck~\cite{Shafi:2010jr, Rehman:2010wm}.  One major difficulty remains, namely what to do with the topological defects that may be copiously produced at the end of inflation (depending on the gauge group chosen).  One option, employed in those earlier treatments, is to specialize to a gauge group which does not result in topological defects, such as the so-called `flipped SU(5)' group (SU(5)$\times$U(1)$_X$).  Fortunately, the main conclusions are upheld under a change in gauge group, so this option may be exploited without detriment to the results.  Another solution is to ensure that the gauge group is broken sufficiently early during inflation so that any cosmologically catastrophic objects are inflated away.  This is the motivation behind the `shifted' class of SUSY hybrid inflation models~\cite{Jeannerot:2000sv}.  In this paper, we will show that a red-tilted spectrum and observable gravity waves, in keeping with the previous treatments, can also be achieved in shifted hybrid inflation while the issue of topological defects is automatically resolved.

The paper is organized as follows.  Section~\ref{sec_BG} contains a brief review of shifted hybrid models, developing the equations that describe the system as well as a synopsis of the differences versus the standard hybrid scenario.  In Section~\ref{sec_min}, we employ a minimal K\"ahler potential and show how this leads to a red-tilted spectrum.  In Section~\ref{sec_nonmin}, we generalize the K\"ahler potential to include non-minimal contributions, and show how this leads to large tensor modes that may be observable by Planck.  Finally, Section~\ref{sec_sum} contains a summary and concluding remarks.


\section{Review of Shifted Hybrid Models}
\label{sec_BG}

\begin{figure}[t]
	\begin{tabular}{cc}
		\includegraphics[width=.5\columnwidth]{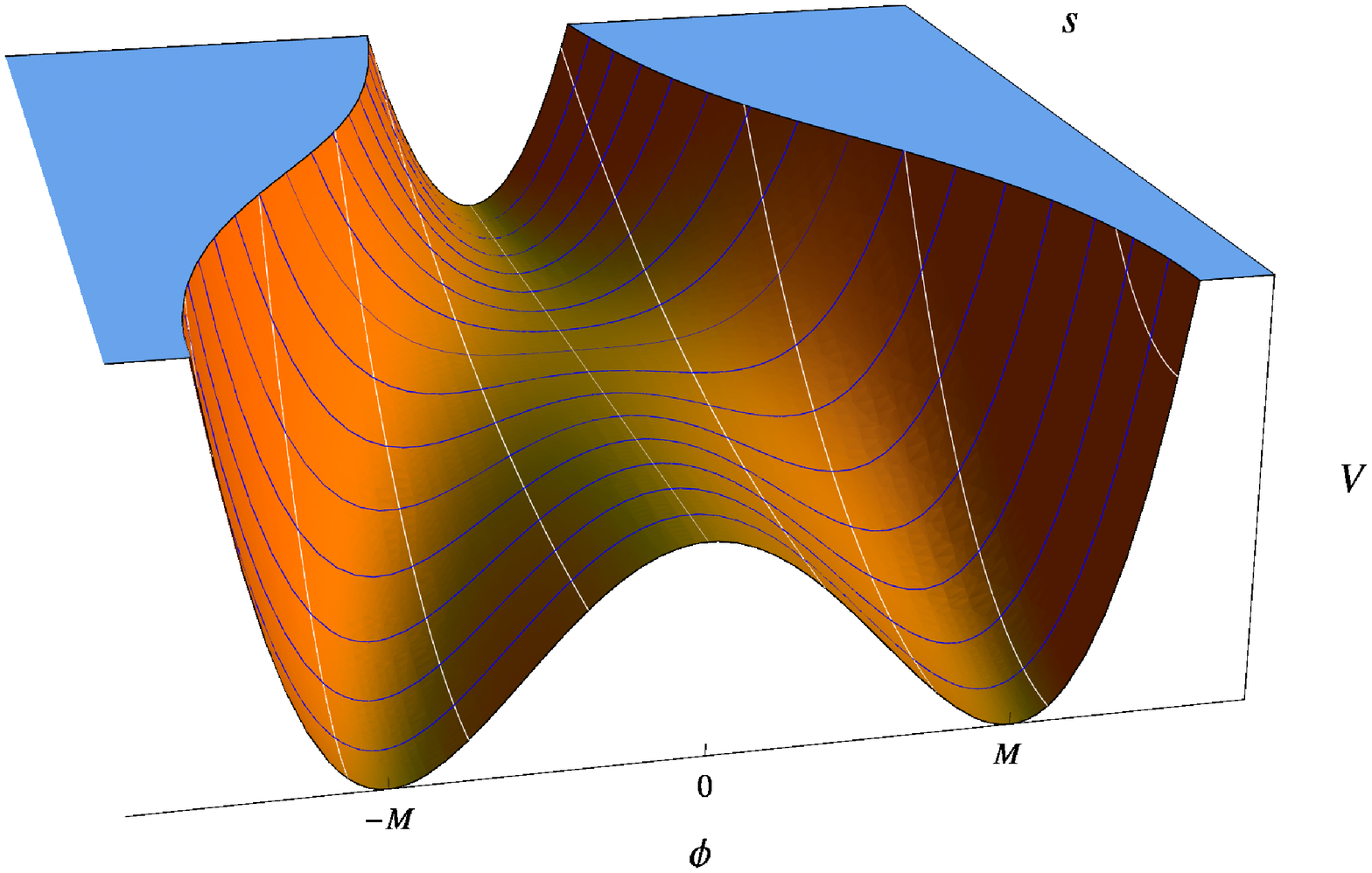} & \includegraphics[width=.5\columnwidth]{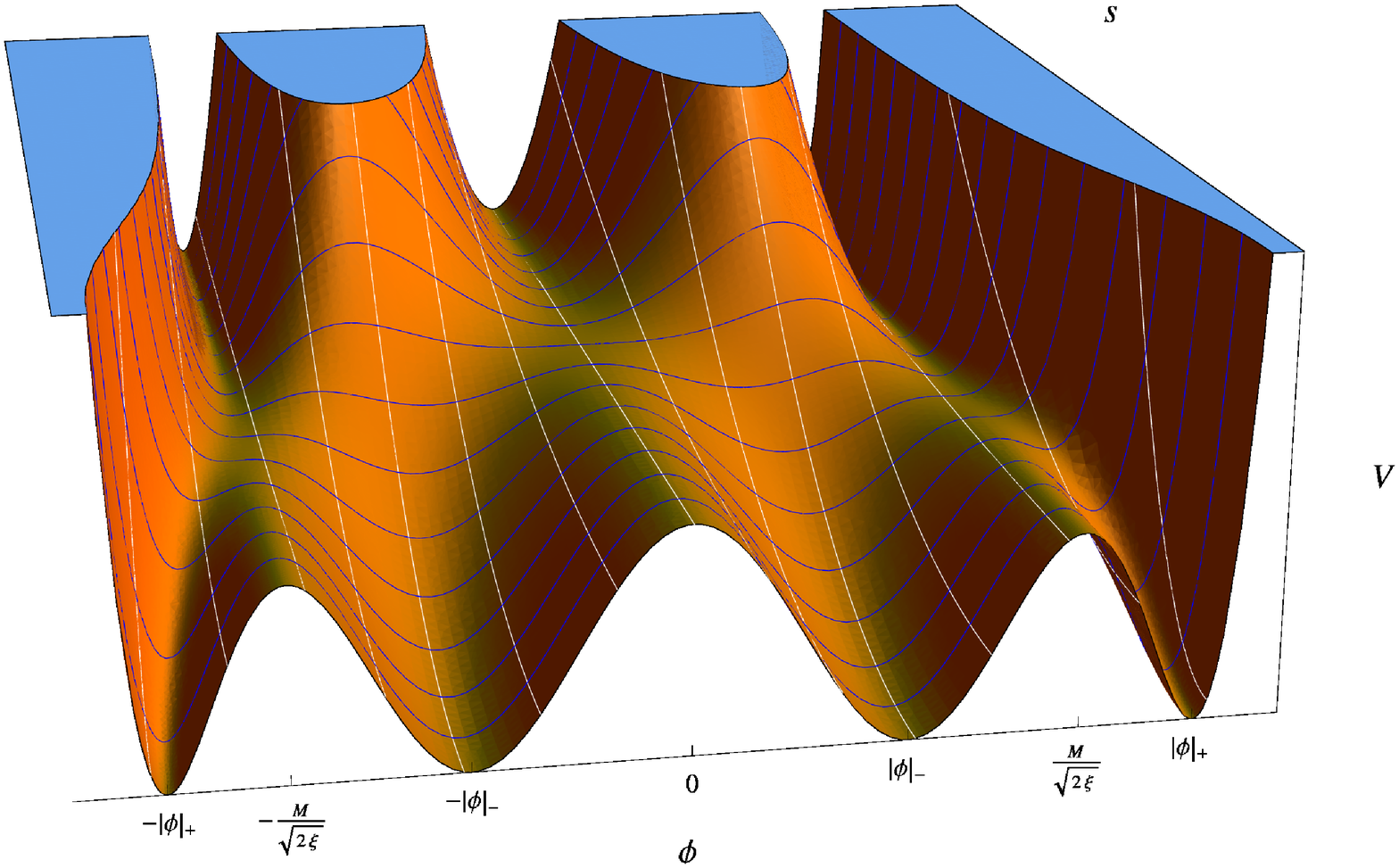} \\
		(a) & (b) 
	\end{tabular}
	\caption{Scalar potentials in the (a) standard and (b) shifted SUSY hybrid inflation scenarios.  Here, we use the potentials derived from the tree level, global SUSY case; additional contributions are important, and are described in the text.  Various locations of interest are marked off in the $\phi$ direction; see Eqs.~(\ref{phiDI}) and (\ref{vev}) for definitions of quantities in the shifted case.}
	\label{3Dplots}
\end{figure}

The simplest model of SUSY hybrid inflation is described by the superpotential~\cite{Dvali:1994ms, Copeland:1994vg}
\begin{equation} 
W_{st} = \kappa S(\overline{\Phi} \Phi - M^{2}) ,
\label{Wstandard}
\end{equation}
where $S$ is a gauge singlet superfield~\cite{*[{For SUSY models inflating with a gauge non-singlet, see }]  [{; }] Antusch:2010va, *Senoguz:2004ky, *Allahverdi:2006iq}, $\Phi$ and $\overline{\Phi}$ are conjugate supermultiplets under some gauge group $G$, $M$ is the energy scale at which $G$ breaks, and $\kappa$ is a dimensionless coupling which we take to be positive without loss of generality.  If a U(1) `$R$-symmetry' is included in the theory, $W_{st}$ is the most general superpotential leaving both $G$ and the $R$-symmetry intact at the renormalizable level.  We will refer to the class of models described by $W_{st}$ as `standard' (SUSY) hybrid inflation.  Panel~(a) of Fig.~\ref{3Dplots} depicts the inflationary scalar potential that is derived from $W_{st}$.


In the standard scenario, there is one flat direction that can support inflation (driven by radiative corrections, and/or other contributions) in the scalar potential $V(s, \phi, \overline{\phi})$ with $s, \phi, \overline{\phi}$ the scalar components of the superfields.  This inflationary valley is aligned along the $s$-direction, with $|\phi| = |\overline{\phi}| = 0$ until the inflaton experiences a `waterfall' at the critical point $|s| \gtrsim M$.  (Note that we take $|\phi| = |\overline{\phi}|$ throughout, in order that the $D$-terms vanish.)  After reaching the waterfall point, inflation ends and the gauge group $G$ is broken, and topological defects may be produced as the system of fields transitions to the vev.

The `shifted' (SUSY) hybrid scenario modifies $W_{st}$ to include an additional non-renormal\-iz\-able term~\cite{Jeannerot:2000sv}~\footnote{If the gauged supermultiplets are in the adjoint representation, the leading order non-renormal\-iz\-able term is cubic in $\Phi$.  For a detailed analysis of shifted inflation with an SU(5) GUT, see Ref.~\cite{Khalil:2010cp}.}
\begin{eqnarray}
W_{sh} &=& \kappa S(\overline{\Phi} \Phi - M^{2}) - \beta S \frac{(\overline{\Phi} \Phi)^2}{M_*^2} , \\
&=& \kappa S \left[ (\overline{\Phi} \Phi - M^{2}) - \xi \frac{(\overline{\Phi} \Phi)^2}{M^2} \right] ,
\label{Wshifted}
\end{eqnarray}
where $\beta$ is a (positive) dimensionless coupling, $M_*$ is some cutoff scale of the theory, and we have defined $\xi \equiv \beta M^2 / \kappa M_*^2$ for convenience.  The (global SUSY) scalar potential in this model appears as
\begin{equation}
V_\text{global} = \kappa^2 \left( \left[ (|\phi|^2 - M^2) - \xi \frac{|\phi|^4}{M^2} \right]^2 + \sigma^2 |\phi|^2 \left[ 1 - 2\xi \frac{|\phi|^2}{M^2}  \right]^2 \right) ,
\label{Vglobal}
\end{equation}
where we have defined the canonically normalized inflaton field $\sigma \equiv \sqrt{2} |s|$, and we have taken the $D$-flat direction with $\arg \phi + \arg \overline{\phi} = 0$.  As can be seen in Panel~(b) of Fig.~\ref{3Dplots}, this potential retains the inflationary track along $|\phi| = 0$, and also contains two additional tracks at constant values
\begin{equation}
|\phi| = \pm \frac{M}{\sqrt{2 \xi}}.
\label{phiDI}
\end{equation}
The shape of the potential along these tracks is very similar to that of the standard track, and so the inflationary dynamics are expected to be much the same as in the standard case.  However, if we choose one of the tracks described by Eq.~(\ref{phiDI}), the inflationary valley is `shifted' to nonzero $|\phi|$ values, and so the gauge symmetry $G$ is broken.  If enough e-foldings of inflation take place along this shifted track, topological defects from $G$-breaking will be inflated away, along with any other objects that are problematic in large densities (e.g. primordial black holes).  In our discussion of shifted hybrid models, we will assume that inflation occurs entirely along a valley described by Eq.~(\ref{phiDI}).\footnote{Inflation along the $|\phi| = 0$ valley within the shifted potential occurs in very much the same way as in the standard hybrid case, the chief difference being the value of $\phi$ at the vev.  In addition, there exist a number of interesting intermediate scenarios where inflation takes place partially along each of the standard and shifted tracks, the details of which are enumerated in Ref.~\cite{Jeannerot:2000sv}; these will not concern us here.}  In this case, we may only have values
\begin{equation}
\frac{1}{8} \leq \xi < \frac{1}{4} ,
\label{xirange}
\end{equation}
in order to ensure that $V_\text{min} = 0$ and that the shifted track lies lower than the standard track (i.e. the appropriate initial conditions are easily achieved)~\cite{Jeannerot:2000sv}.  Additionally, we will assume that $G$ is broken down to the Standard Model in a single stage, so that multiple periods of inflation are not needed to handle topological defects produced by subsequent stages of symmetry breaking.

In order to drive inflation, a slope must be given to the flat direction(s) in the potential $V_\text{global}$ above.  Eq.~(\ref{Vglobal}) is a tree level expression, and since SUSY is broken (i.e. $V > 0$ due to $F_S \neq 0$) along this flat direction, there is a mass splitting in the supermultiplets and the radiative corrections to the potential do not vanish entirely.  Using the Coleman-Weinberg formula~\cite{Coleman:1973jx}, the 1-loop contribution may be written as~\cite{Jeannerot:2000sv}
\begin{eqnarray}
\Delta V_\text{1-loop} &=& \kappa^2 m^4 \cdot \frac{\kappa^2}{4 \pi^2} F(x) , \\
F(x) &=& \frac{1}{4} \left[ (x^4 + 1) \ln\frac{(x^4 - 1)}{x^4} + 2 x^2 \ln\frac{x^2 + 1}{x^2 - 1} + 2 \ln\left( 2 \frac{\kappa^2 m^2 x^2}{Q^2} \right) - 3 \right] ,
\label{Vloop}
\end{eqnarray}
where $Q$ is the renormalization scale, and we have defined the useful parametrizations
\begin{eqnarray*}
m^2 &\equiv& M^2 \left( \frac{1}{4 \xi} - 1 \right) , \\
x &\equiv& \frac{\sigma}{m} ,
\end{eqnarray*}
which we will use throughout our analysis.  Once the potential is given a slope, the inflaton rolls toward smaller field values, until its instantaneous mass becomes tachyonic at $x = 1$ ($\sigma = m$).  The system becomes destabilized and undergoes a `waterfall' transition to the vev.  The vacuum appears at $\sigma = 0$, but in the $\phi$ direction there exist two vacua on each side of the origin at values
\begin{equation}
|\phi|_\pm^2 \xrightarrow{\sigma = 0} \frac{M^2}{2 \xi} \left[ 1 \pm \sqrt{1 - 4 \xi} \right] .
\label{vev}
\end{equation}
As described in Ref.~\cite{Jeannerot:2000sv}, $|\phi|_-$ appears earlier than $|\phi|_+$ as the inflaton rolls, and the system evolves into $|\phi|_-$ before $|\phi|_+$ exists as a minimum.  Hence the appropriate choice of global vev is $\langle |\phi| \rangle = |\phi|_-$.

In general, there exist further contributions to the potential in addition to radiative corrections.  Although we have already noted that SUSY is spontaneously broken during inflation, we may also have contributions from explicit soft SUSY-breaking terms.  We may write the effective linear and mass-squared soft terms in the form
\begin{eqnarray}
\Delta V_\text{soft} &=& \frac{1}{\sqrt{2}} a m_{3/2} \kappa m^3 x + \frac{1}{2} M_\sigma^2 m^2 x^2 + M_\phi^2 \left( \frac{M^2}{\xi} \right) , \\
a &=& 2 |A - 2| \cos [\arg S + \arg (A-2)],
\label{Vsoft}
\end{eqnarray}
with $m_{3/2} \sim 1$~TeV, and where $A - 2$ is the complex coefficient of the linear soft term in the Lagrangian.  It has been shown that $a$ does not vary much over the course of inflation if $\arg S$ is initially very small~\cite{urRehman:2006hu}.  We will tacitly assume that this choice has been made, and treat $a$ as a constant of order unity.  The soft masses $M_\sigma$ and $M_\phi$ can, in principle, lie at intermediate scales; indeed, it has been shown that such a choice can lead to a favorable reduction in $n_s$ (for $M_S^2 < 0$)~\cite{Rehman:2009yj} or to Planck-observable values of $r$ (for $M_S^2 > 0$)~\cite{Shafi:2010jr} in the context of the standard hybrid scenario.  However, it has also been shown that these intermediate scales are not crucial~\cite{Rehman:2009nq, Rehman:2010wm} to either effect.  It is perhaps more attractive from a model building perspective to employ soft masses around the TeV-scale, so we will take $M_\sigma, M_\phi \sim m_{3/2}$.

Finally, it is inconsistent to use global SUSY to describe physics near the Planck scale.  Rather, we should employ supergravity.  The $F$-term SUGRA scalar potential is a function of both the superpotential $W$ and the K\"ahler potential $K$ of a given theory, and may be written down using the formula
\begin{equation} 
V_{F} = e^{K/m_{P}^{2}} \left( K_{ij}^{-1}D_{z_{i}}WD_{z^{*}_j}W^{*} - 3m_{P}^{-2}\left| W\right| ^{2} \right),
\label{VF1}
\end{equation}
where $z_{i}\in \{s, \phi , \overline{\phi }, \cdots\}$, and we have used the shorthand
\begin{eqnarray*}
K_{ij} &\equiv& \frac{\partial ^{2}K}{\partial z_{i}\partial z_{j}^{*}}, \\
D_{z_{i}}W &\equiv& \frac{\partial W}{\partial z_{i}}+m_{P}^{-2}\frac{\partial K}{\partial z_{i}}W, \\ 
D_{z_{i}^{*}}W^{*} &=& \left( D_{z_{i}}W\right) ^{*}.
\end{eqnarray*}

We will have occasion to use both minimal (canonical) and non-minimal forms of the K\"ahler potential.  In general, $K$ should be expanded in inverse powers of the cutoff scale $M_*$
\begin{multline}
K = |S|^2 + |\Phi|^2 + |\overline{\Phi}|^2 + \frac{\kappa_S}{4}\frac{|S|^4}{M_*^2} + \frac{\kappa_\Phi}{4}\frac{|\Phi|^4}{M_*^2} +\frac{\kappa_{\overline{\Phi}}}{4}\frac{|\overline{\Phi}|^4}{M_*^2} \\
 + \kappa_{S \Phi}\frac{|S|^2|\Phi|^2}{M_*^2} + \kappa_{S \overline{\Phi}}\frac{|S|^2|\overline{\Phi}|^2}{M_*^2} + \kappa_{\Phi \overline{\Phi}}\frac{|\Phi|^2|\overline{\Phi}|^2}{M_*^2} + \frac{\kappa_{SS}}{6}\frac{|S|^6}{M_*^4} + \cdots .
\label{kahler}
\end{multline}
However, for reasons that will become clear later, we will choose $M_* = m_P$ in the cases where we have need for the higher-order (i.e. cutoff suppressed) terms in $K$.  (In the cases where we use the minimal K\"ahler potential, $M_*$ will enter only via the definition of $\xi$.)  Under this assumption, the SUGRA terms in the scalar potential will be suppressed by the Planck scale, and we will keep only the lowest order (quadratic and quartic) contributions from this source.

\begin{figure}[t]
	\includegraphics[width=.8\columnwidth]{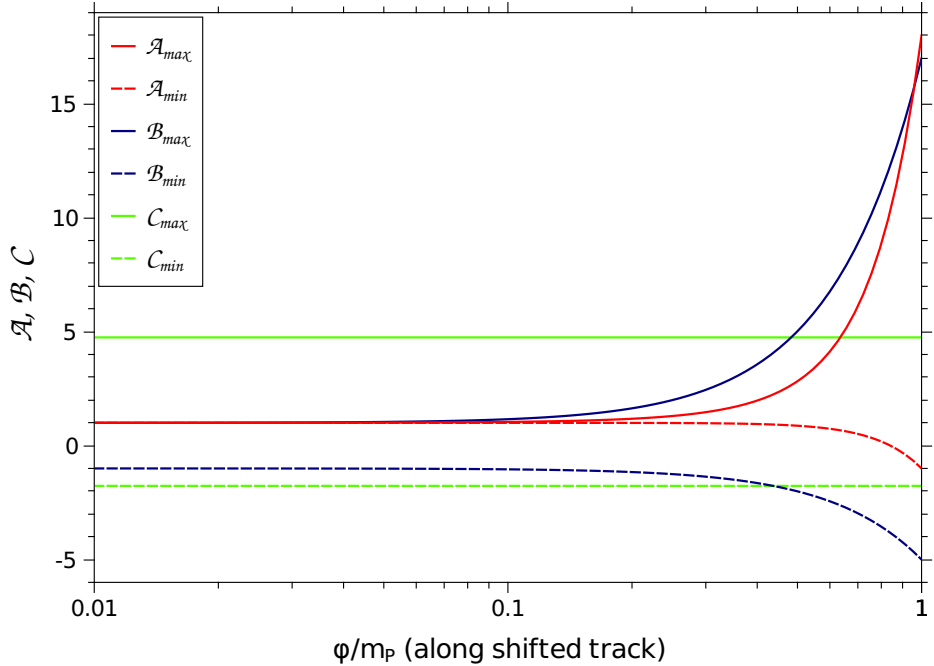}
	\caption{The upper and lower bounding curves for the effective couplings $\mathcal{A}$ (red), $\mathcal{B}$ (blue), and $\mathcal{C}$ (green).  The solid lines signify the upper bounding function for each parameter according to Eqs.~(\ref{Amax})--(\ref{Cmin}), while the dashed curves represent the lower bounding function.  These parameters may be further constrained by other considerations, as explained in the text.}
	\label{coefflimits}
\end{figure}

Putting this all together, we may write the inflationary potential as
\begin{eqnarray}
V &=& V_\text{global}(|\phi| = \frac{M}{\sqrt{2 \xi}}) + \Delta V_\text{SUGRA} + \Delta V_\text{1-loop} + \Delta V_\text{soft} , \nonumber \\
  &=& \kappa^2 m^4 \left[ \mathcal{A} + \frac{1}{2} \mathcal{B} \left(\frac{m}{m_P}\right)^2 x^2 + \frac{1}{4} \mathcal{C} \left(\frac{m}{m_P}\right)^4 x^4  +  \frac{\kappa^2}{4 \pi^2} F(x) \right] \nonumber \\
  && \quad\quad + \frac{1}{\sqrt{2}} a m_{3/2} \kappa m^3 x + \frac{1}{2} M_\sigma^2 m^2 x^2 + M_\phi^2 \left( \frac{M^2}{\xi} \right) .
\label{Vfull}
\end{eqnarray}
The effective coefficients $\mathcal{A}, \mathcal{B}, \mathcal{C}$ are complicated functions of the couplings $\kappa_i$ in the K\"ahler potential, and of the quantity $\phi_P \equiv |\phi|/m_P = (M/m_P)/\sqrt{2 \xi}$~(for explicit expressions, see Appendix~\ref{app}).  We will find it convenient to work directly with these parameters, since many different arrangements of the values of the $\kappa_i$ couplings can lead to degenerate results.  If we take all the non-minimal couplings in the natural range $-1 \lesssim \kappa_i \lesssim 1$, we obtain the following extremal functions:
\begin{eqnarray}
\mathcal{A}_\text{max} &=& 1 + 4 \phi_P^2 + 13 \phi_P^4 , \label{Amax} \\
\mathcal{A}_\text{min} &=& 1 - 2 \phi_P^4 , \\
\mathcal{B}_\text{max} &=& 1 + 16 \phi_P^2 , \\
\mathcal{B}_\text{min} &=& -1 - 4 \phi_P^2 , \\
\mathcal{C}_\text{max} &=& \frac{19}{4} , \\
\mathcal{C}_\text{min} &=& -\frac{113}{64} . \label{Cmin}
\end{eqnarray}
The behavior of these functions is depicted in Fig.~\ref{coefflimits}.  Note that the vacuum potential is now $V_0 = \kappa^2 m^4 \mathcal{A}$, so we require that $\mathcal{A} \gtrsim 0$.  In order that perturbativity be preserved, we also enforce $|\mathcal{B}|, |\mathcal{C}| < \mathcal{A}$.


Having written down the potential, we may now calculate predictions for observable quantities via the usual slow roll formulation.  The slow roll parameters relevant to our discussion are written in terms of $x$ as
\begin{eqnarray}
\epsilon &=& \frac{1}{2} \left(\frac{m_P}{m}\right)^2 \left( \frac{V'}{V}\right) ^{2}, \label{epsilon} \\
\eta &=& \left(\frac{m_P}{m}\right)^2 \left( \frac{V''}{V}\right), \label{eta} 
\end{eqnarray}
where primes denote differentiation with respect to $x$.  The slow-roll approximation corresponds to $\epsilon$, $|\eta| \ll 1$, and the duration of inflation is parametrized in terms of the number of e-foldings
\begin{equation}
N_{0}=\left( \frac{m}{m_{P}}\right) ^{2}\int_{x_e}^{x_{0}}\left( \frac{V}{V'}\right) dx ,
\label{N0}
\end{equation}
where a subscript `0' denotes a value taken when the pivot scale $k_0 = 0.002 \text{ Mpc}^{-1}$ crosses the horizon.  In principle, the value $x_e$ at the end of inflation is specified by the waterfall transition at the value $x_c = 1$.  In practice, the slow roll parameter $\eta$ blows up in the limit $x \rightarrow x_c$; inflation ends when $|\eta| \sim 1$, which may occur very close to $x = x_c$.  It is also possible for the slow roll approximation to break down earlier along the inflationary trajectory, in which case the value $x_e$ is fixed at the field value where this occurs.  To leading order in the slow roll parameters, the scalar spectral index, tensor-to-scalar ratio, and primordial curvature perturbation appear as
\begin{eqnarray}
n_s &\simeq& 1 - 6\epsilon + 2\eta, \label{nsSR} \\
r &\simeq& 16\epsilon, \label{rSR} \\
\Delta_{\mathcal{R}}^2 &\simeq& \frac{m^2}{12 \pi^2 m_{P}^{6}}\left( \frac{V^{3}}{(V')^2}\right). \label{curvSR}
\end{eqnarray}
These quantities will be evaluated at the pivot scale $x_0$, in order to compare with the experimental measurements from WMAP7~\citep{Komatsu:2010fb}.

From Eq.~(\ref{Vfull}), the similarities between this shifted model and the standard hybrid scenario are now manifest.  Indeed, by comparing potentials, it is a straightforward matter to transform one model into the other by a suitable redefinition of various parameters.  The most important difference is that the mass parameter $m$ in the shifted model has taken the place of $M$ in the vacuum potential, and also in specifying the location of the waterfall.  Owing to Eq.~(\ref{xirange}), we have $m \leq M$, so this will tend to reduce the vacuum potential relative to the standard case.\footnote{Note, however, that for successful inflation to occur, we should not see the vacuum potential deviating too much from the values obtained in the standard hybrid scenario.}  On the other hand, if $\mathcal{A}$ is allowed to be significantly greater than unity, we will also have an enhancement in $V_0$.  The size of $\mathcal{A}_\text{max}$ is limited by the size of $\phi_P$, which in turn is limited by $M$.  But $M$ should not be too close to the Planck scale for reasons of consistency; if we take $M \lesssim 0.1 m_P$, we have $\phi_P \lesssim 0.2$, and we see from Fig.~\ref{coefflimits} that the system is confined in a region where $\mathcal{A} \sim 1$ is forced upon us.

In addition, we pick up small changes to the contributions from radiative and soft corrections.  The coefficient of the radiative correction function contains an extra factor of 2 relative to the standard hybrid model.  Also, the factor $\mathcal{N}$ corresponding to the size of the gauge representation of $\phi, \overline{\phi}$ in the standard case is missing in the shifted case; this is because the gauge symmetry $G$ is broken, and we can no longer describe fields in terms of $G$-multiplets along the shifted track.  Finally, the factor 2 inside the last logarithm in $F(x)$ is absent in the standard case, as a consequence of defining $x$ in terms of $|s|$ rather than $\sigma$.  The primary difference in the soft terms is that the soft mass-squared term for $\phi, \overline{\phi}$ is nonzero in shifted inflation.  However, since we take soft masses of order $\sim 1$~TeV, we do not expect this to have a substantial effect on the results.

In the calculations discussed in the following sections, we will make a handful of broad, mostly general assumptions.  We will take all mass scales no larger than $m_P$, usually somewhat smaller, to ensure that series expansions remain perturbative and that quantum gravity effects do not become important.  We will also assume that all physical couplings are at most unity; in particular, this will involve placing a manual constraint $\beta \leq 1$, since we choose the alternative $\xi$ as an independent variable.  Finally, it can be shown that the inflaton mass-squared $m_\text{inf}^2$ at the vev remains positive for the full range of parameters we will use.


\section{Shifted Inflation and the Spectral Index}
\label{sec_min}

In the case of minimal K\"ahler, we have
\begin{eqnarray}
\mathcal{A} &\rightarrow& 1 + 2 \left[ \phi_P^2 + \phi_P^4 \right] , \label{Acan} \\
\mathcal{B} &\rightarrow& 2 \phi_P^2 , \label{Bcan} \\
\mathcal{C} &\rightarrow& \frac{1}{2} . \label{Ccan}
\end{eqnarray}
As mentioned earlier, the cutoff $M_*$ only appears in the definition of $\xi$ here, as does $\beta$.  From general considerations, we expect to have $M_* \lesssim m_P$ and $\beta \lesssim 1$.  If we fix one of these parameters, the other may be calculated for given values of $\kappa$, $M$ and $\xi$, each of which will be varied independently.  If we fix $M_* \simeq m_P$, it turns out that $\beta$ should be large ($1 \lesssim \beta \lesssim 100$) so that $\xi$ does not grow too small.  Then, in this version of the model, we will fix $\beta \simeq 1$, which leads to cutoff values below the Planck scale.

With Ref.~\cite{Rehman:2009nq} as our inspiration, we choose $a = -1$ for the coefficient of the linear soft term, and hold the soft masses at the TeV scale.  In this case, we expect to find regions of parameter space where the spectral index $n_s$ can be smaller than 1 in accordance with the latest WMAP data.  For comparison, choosing $a = 0, +1$ leads to $n_s \gtrsim 0.985$, lying just outside the WMAP 1$\sigma$ bound.

\begin{figure}[pt]
	\begin{tabular}{cc}
		\includegraphics[width=.5\columnwidth]{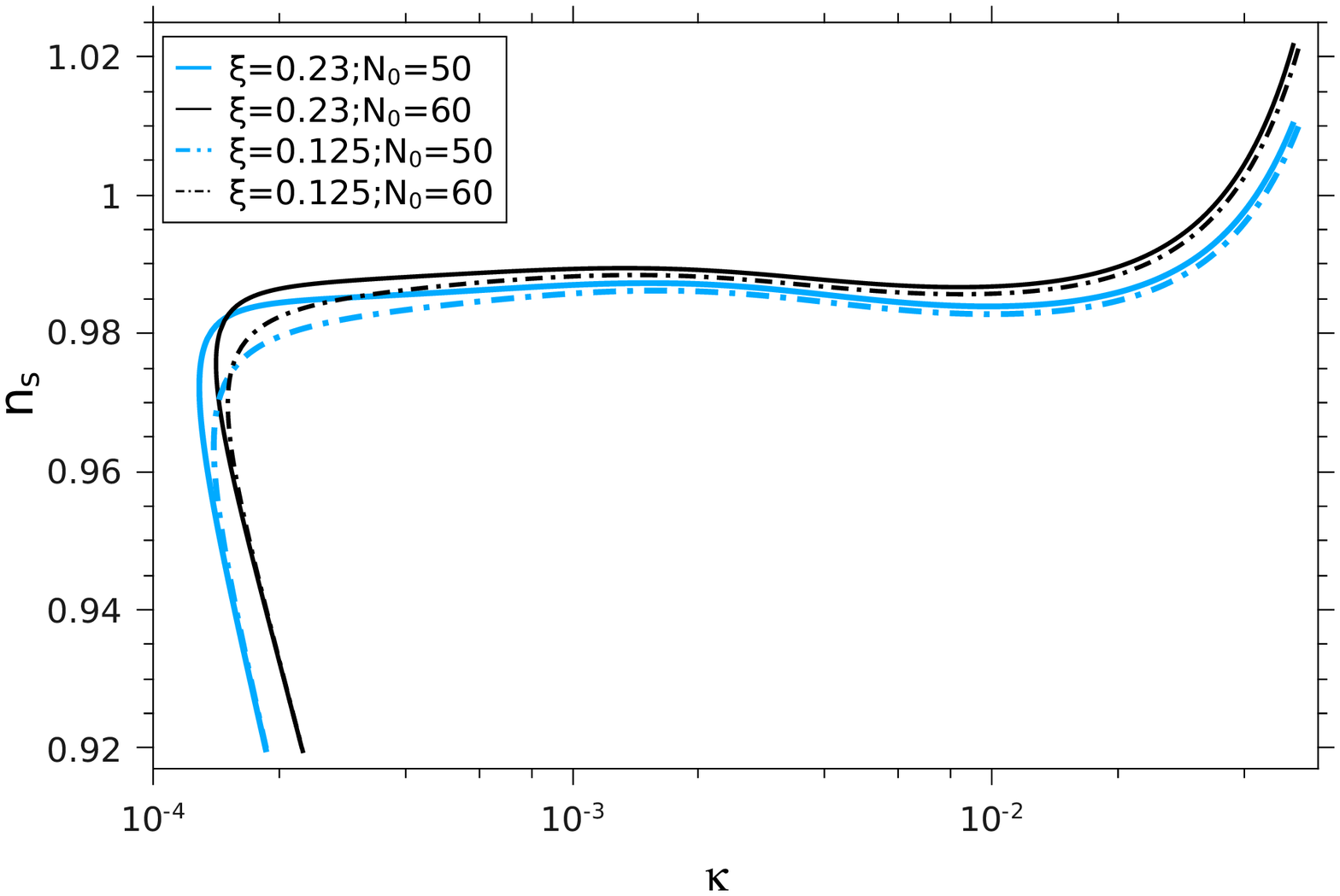} & \includegraphics[width=.5\columnwidth]{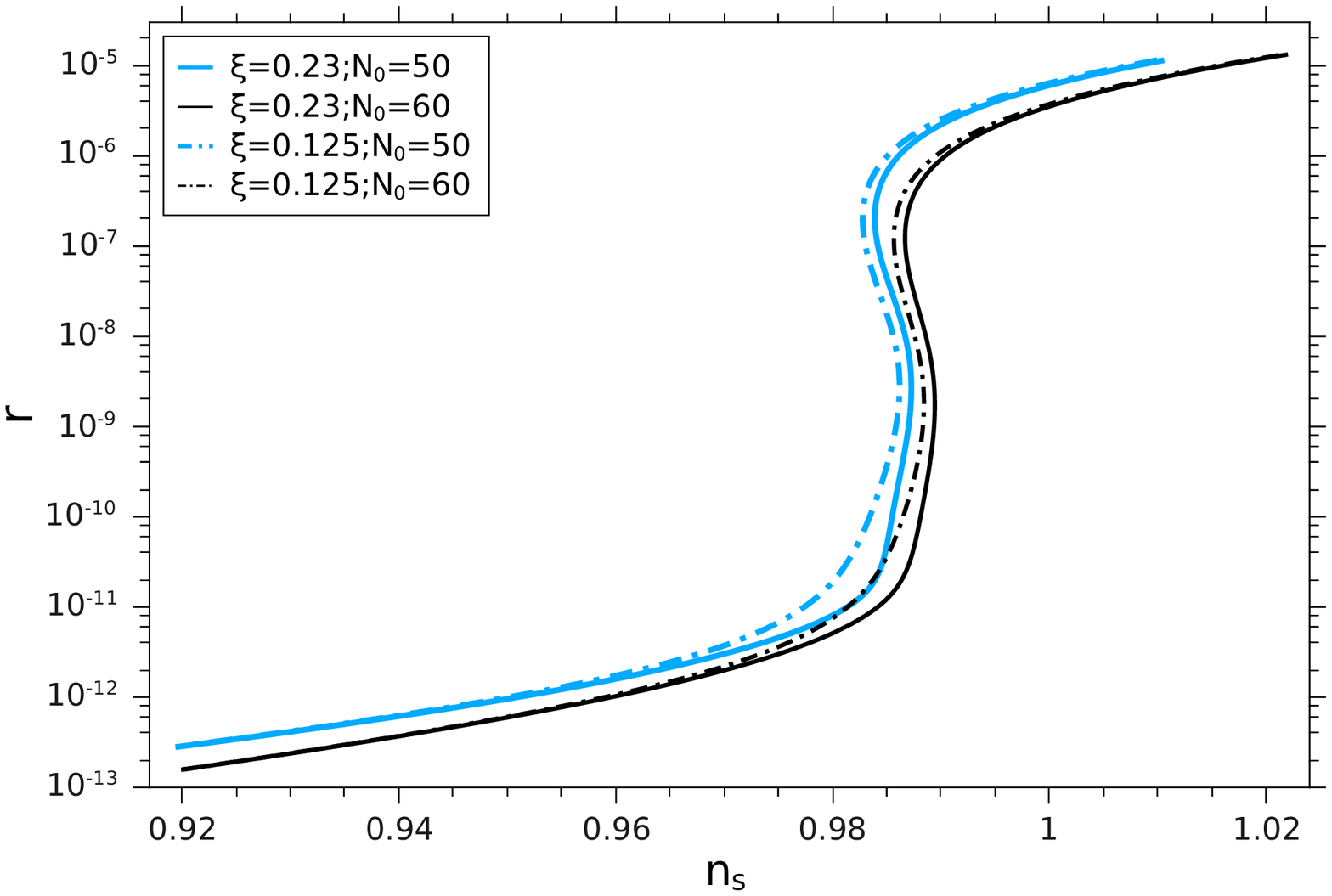} \\
		(a) & (b) 
	\end{tabular}
	\includegraphics[width=.5\columnwidth]{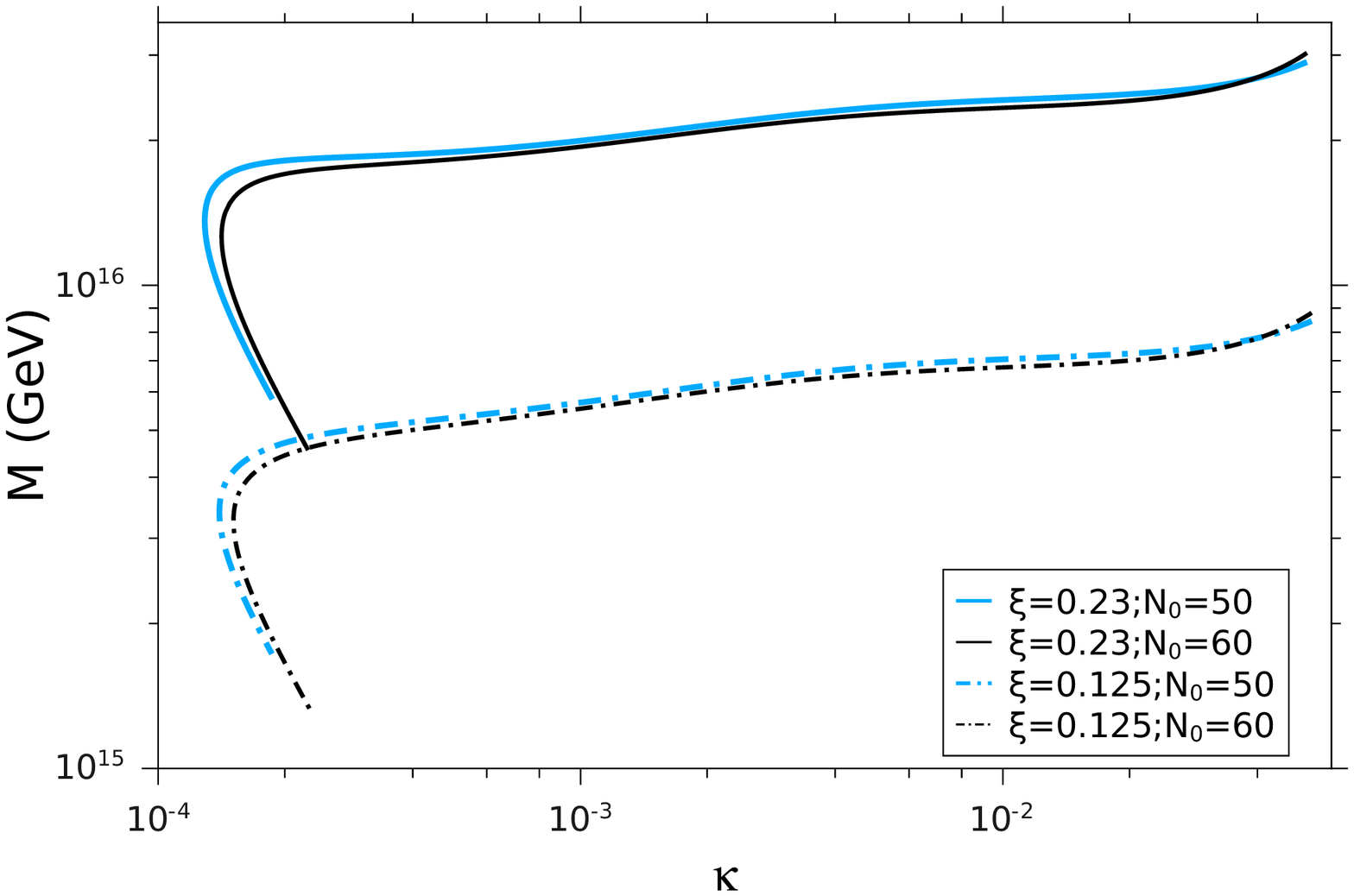} \\
	(c)
	\caption{Results of our numerical calculations for the shifted hybrid inflation model with a minimal (i.e. canonical) K\"ahler potential.  Here, we have taken fixed values of $\xi$ and $N_0$ at/near the endpoints of their ranges, and intermediate results can be interpolated between the curves.  As described in the legend for each panel, the light (dark) lines represent 50 (60) e-foldings, whereas the line pattern specifies the fixed value of $\xi$.}
	\label{min_results}
\end{figure}

The results of the calculations for the minimal K\"ahler case are displayed in Panels~(a)--(c) of Fig.~\ref{min_results}.  In these plots, we fix $\xi$ and the number of e-foldings $N_0$ at the extremities of their ranges (with $50 \lesssim N_0 \lesssim 60$) for each curve, and a family of curves can be interpolated to sweep out the allowed region.  The behavior exhibited here is indeed very similar to the analogous results from the standard hybrid case.  While there exists a sizable shift in $M$ for different values of $\xi$, recall that $m$ is playing the same dynamical role in the shifted case as that of $M$ in the standard case, and so we expect such a shift.


These results are also quantitatively similar to the standard case, but some subtle shifts and changes in curve shapes are exhibited.  This can be understood by noting that additional terms exist in the present case.  Using Eqs.~(\ref{Acan})--(\ref{Ccan}), it is clear that the values needed to recover the standard hybrid case (i.e. $\mathcal{A} = 1$, $\mathcal{B} = 0$) are only produced up to some extra Planck-suppressed contributions.  In addition, there are some small differences in the contribution coming from radiative corrections as previously noted.

Similarly to the standard hybrid case, the use of a negative soft term has enabled the model to achieve a red-tilted spectrum in agreement with WMAP7, but the tensor-to-scalar ratio $r$ is quite small (particularly in the region where $n_s$ is most favored).  The existence of tensor modes in the Cosmic Microwave Background (CMB) is currently being tested by the Planck Satellite observatory, and can serve as a discriminatory measure between similar models of inflation.  Indeed, if a precise measurement of $r$ is made by Planck, we will obtain information on the energy scale of inflation, and if Planck merely sets more stringent bounds many inflation models may be ruled out.  Thus the question of whether the current model may, under appropriate circumstances, predict large tensor modes is one worth careful consideration.


\section{Shifted Inflation and Gravity Waves}
\label{sec_nonmin}

Given the similarities to the standard hybrid inflation model already noted, we may look to that case for an idea of what to expect in the shifted hybrid inflation model.  In Refs.~\cite{Shafi:2010jr, Rehman:2010wm}, we have examined the case in which SUGRA corrections using a non-minimal K\"ahler potential are included in the standard hybrid potential.  In that case, two extra parameters are included in the system versus the minimal case, and the additional freedom leads to the possibility of large tensor modes.  The preferred circumstances for the large-$r$ regime include a potential with a $(+ \text{quadratic} - \text{quartic})$ structure, with a negative second derivative (which also turns out to be favorable for producing $n_s < 1$, as in hilltop inflation models~\cite{Boubekeur:2005zm, *[{For hilltop inflation in the context of SUGRA hybrid inflation, see }]  [{.}] Pallis:2009pq}).  In this case, the quartic coefficient needs to be small enough that the quadratic term still dominates in the region where inflation can take place, so that the potential remains stable in the regime of physical importance.  We expect these properties to also be exhibited in the shifted model; although we note that the additional freedom introduced by the non-minimal K\"ahler terms will involve more than just two extra parameters, we consolidate their contributions into $\mathcal{A}, \mathcal{B}$ and $\mathcal{C}$.

As in the standard hybrid scenario, we make one nontrivial assumption in order to eliminate the possibility of a complication.  Under certain circumstances, the potential in Eq.~(\ref{Vfull}) can contain a metastable vacuum along the inflationary track.  If the inflaton becomes trapped in such a false vacuum state, inflation can last for a very long time, ending only if the system escapes the minimum either classically or via tunneling (either of which may require a tremendous amount of energy).  We are interested only in the situation where successful inflation may occur, ending in oscillations about the true (SUSY) vacuum to reheat the universe.  Thus we place a constraint on the potential to ensure that no metastable vacua appear.  We do this by requiring that the potential is essentially monotonic along the entire inflationary trajectory, i.e. between $x_0$ and $x_e$.  In hybrid inflation models, the inflaton rolls from large to small values, so we take the potential to be monotonically increasing ($V' \gtrsim 0$).

Before delving into the full numerical calculation, it is worth pausing a moment to consider the behavior of $r$ at its largest values via analytical approximation.  Large values of $r$ occur for the inflaton close to the Planck scale.  It is convenient to define $f \equiv \sigma / m_P$, which we expect to tend toward unity as $r$ increases.  Then, the polynomial terms in the potential will dominate unless $\kappa$ is quite large, and we may write an effective form of the potential as
\begin{equation*}
V \simeq V_0 \left[ 1 + \frac{1}{2} \widetilde{\mathcal{B}} f^2 + \frac{1}{4} \widetilde{\mathcal{C}} f^4 \right] ,
\end{equation*}
where $\widetilde{\mathcal{B}} \equiv \mathcal{B}/\mathcal{A}$, and $\widetilde{\mathcal{C}}$ is defined similarly.  Approximating $V(x_0) \approx V_0 = \kappa^2 m^4 \mathcal{A}$ in Eqs.~(\ref{epsilon}) and~(\ref{eta}), and using these expressions in Eq.~(\ref{rSR}), we obtain
\begin{equation}
r \simeq 8 f^2 \left[ \widetilde{\mathcal{B}} + \widetilde{\mathcal{C}} f^2 \right]^2 ,
\end{equation}
which can be rewritten in the alternate forms
\begin{eqnarray*}
r &\simeq& 8 f^2 \left[ -2 \widetilde{\mathcal{C}} f^2 + \eta \right]^2 , \\
r &\simeq& \frac{8}{9} f^2 \left[ 2 \widetilde{\mathcal{B}} + \eta \right]^2 .
\end{eqnarray*}
These expressions are useful in predicting the interdependence between $r$ and various parameters in the region where $r$ is largest.  For example, since we have taken $V' \gtrsim 0$ during inflation, we are led to the conclusion that $\widetilde{\mathcal{C}} \sim \mathcal{C}$ should be negative and tending toward zero as $r$ approaches its largest values~\cite{Shafi:2010jr}.  If we desire a more useful (albeit less compact) expression in terms of $n_s$, for which we may employ measured values, we may substitute for $\eta$ using Eq.~(\ref{nsSR}).  This is easily done, but the result is less than enlightening and we will not write it here.

We have obtained an approximate expression for (large) $r$, which is reasonably accurate and useful in a variety of ways, in the regime where the radiative corrections may be reliably suppressed.  However, if $\kappa$ is sufficiently large, this contribution must be taken into account.  As described in detail in Ref.~\cite{Rehman:2010wm}, the radiative corrections play a critical role in placing an upper bound on the value of $r$ in these models.  This happens indirectly via the number of e-foldings $N_0$; as $\kappa$ increases from small values, $r$ increases until the radiative correction term becomes comparable to the polynomial terms in the integrand of Eq.~(\ref{N0}), then begins to decrease.  Thus $r$ cannot be arbitrarily large in these models.  Interestingly enough, as we will see, the upper limit on $r$ essentially corresponds to the smallest values to which the Planck Satellite is expected to be sensitive, and so it will be exciting to see what Planck will have to say about the validity of these models in the near future.


\begin{table}[bt]
\begin{tabular}{c|c|c||c|c}
{\bf Fundamental} & {\bf Range} & {\bf Scale} & {\bf Derived} & {\bf Constraining range} \\
{\bf parameter} &  & {\bf type} & {\bf quantity} &  \\
\hline
$\kappa$ & $[10^{-4}, 5]$ & log & $n_s$ & $[0.920, 1.016]$ \\
$M/m_P$ & $[10^{-4}, 10^{-1}]$ & log &  & $ = 0.968 \pm 4\sigma$ \\		
\cline{4-5}
$\xi$ & $[\frac{1}{8}, \frac{1}{4})$ & linear & $\Delta_{\mathcal{R}}^2$ & $[2.21, 2.65] \times 10^{-9}$ \\      
$\mathcal{A}$ & $[\mathcal{A}_\text{min}, \mathcal{A}_\text{max}]$ & linear &  & $= 2.43 \times 10^{-9} \pm 2\sigma$ \\      
\cline{4-5}
$\mathcal{B}$ & $[10^{-6}, min(\mathcal{B}_\text{max}, \mathcal{A})]$ & log & $r$ & $< 1$ \\
$\mathcal{C}$ & $[max(\mathcal{C}_\text{min}, -\mathcal{A}), min(\mathcal{C}_\text{max}, \mathcal{A})]$ & linear & $N_0$ & $[50,60]$ \\
$a$ & $\lbrace -1,0,1 \rbrace$ & --- &  &  \\
$x_0$ & $[1, \frac{m_P}{m}]$ & linear &  &  \\

\end{tabular}
\caption{Ranges specified for the fundamental parameters in Eq.~(\ref{Vfull}), and constraints placed manually on derived quantities.  Note that $a$ was considered at discrete values, and $x_0$ can take on any value between the waterfall point and the Planck scale.  Central values and standard deviations for measured quantities are in reference to the WMAP 7-year analysis~\cite{Komatsu:2010fb}.  It should be noted that the constraint placed manually on $r$ is designed only to eliminate spurious results, and is not related to the bound given by WMAP7.}
\label{rangetable}
\end{table}


\begin{figure}[pt]
	\begin{tabular}{cc}
		\includegraphics[width=.5\columnwidth]{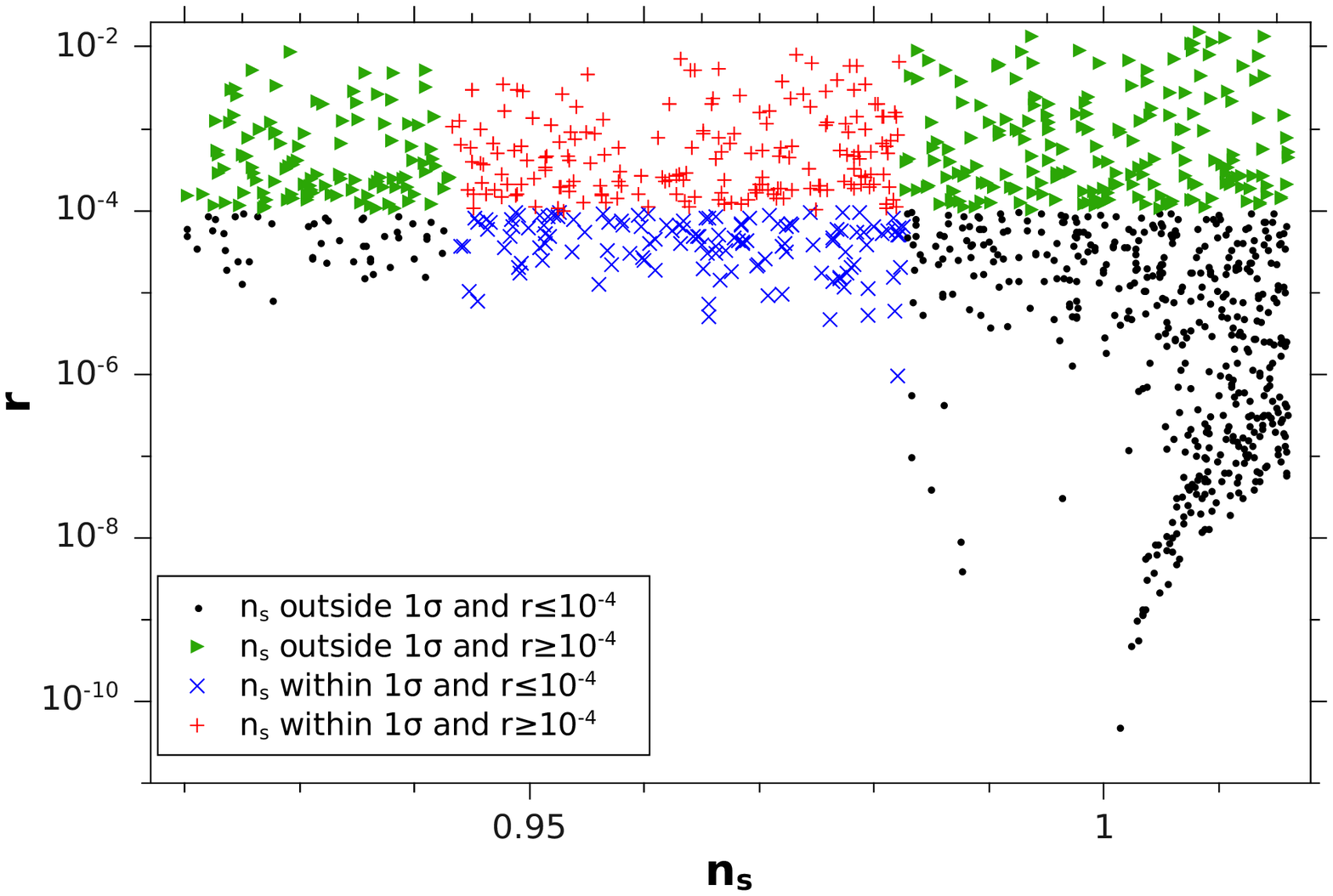} & \includegraphics[width=.5\columnwidth]{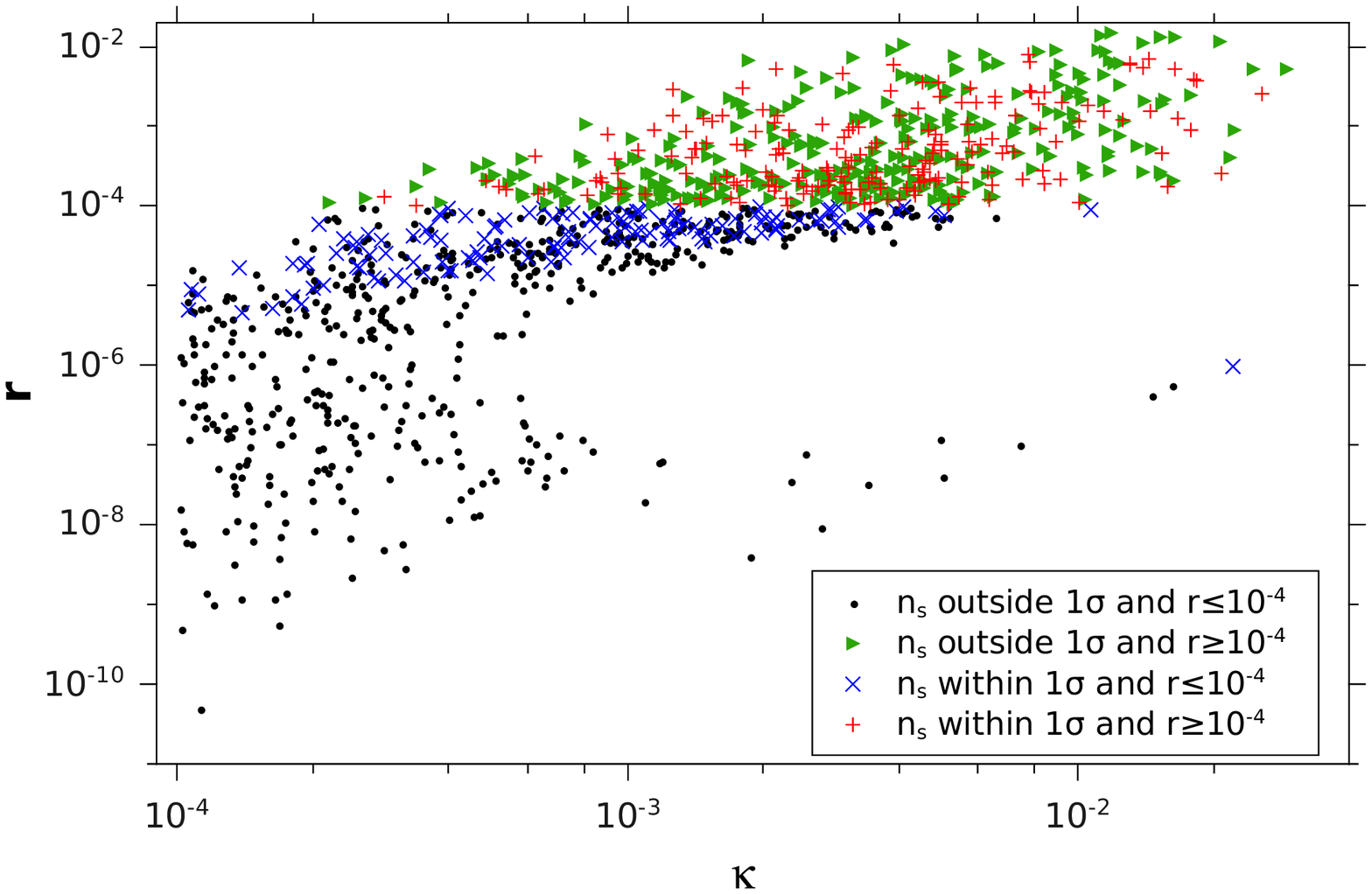} \\
		(a) & (b) \\
		\includegraphics[width=.5\columnwidth]{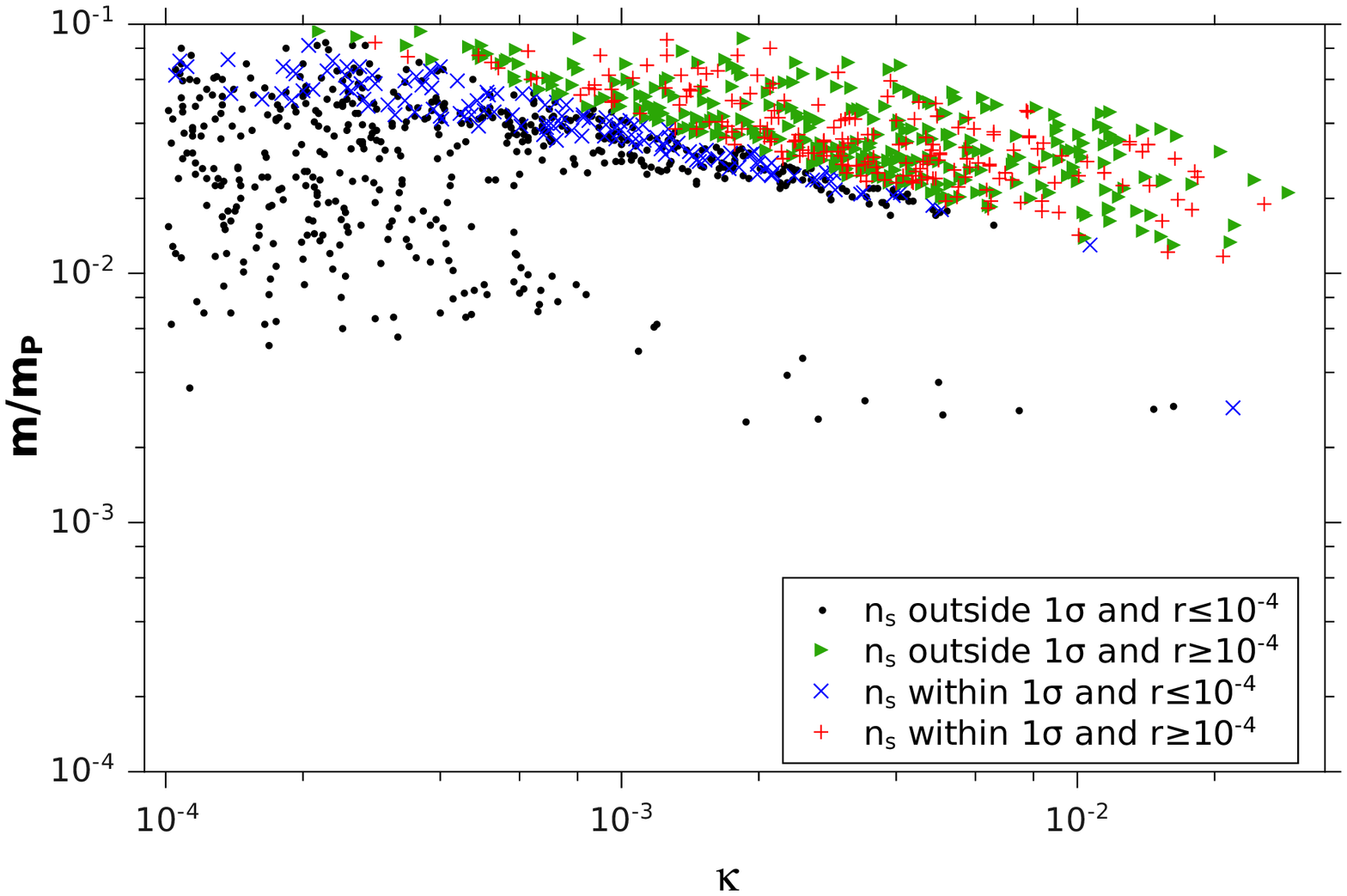} & \includegraphics[width=.5\columnwidth]{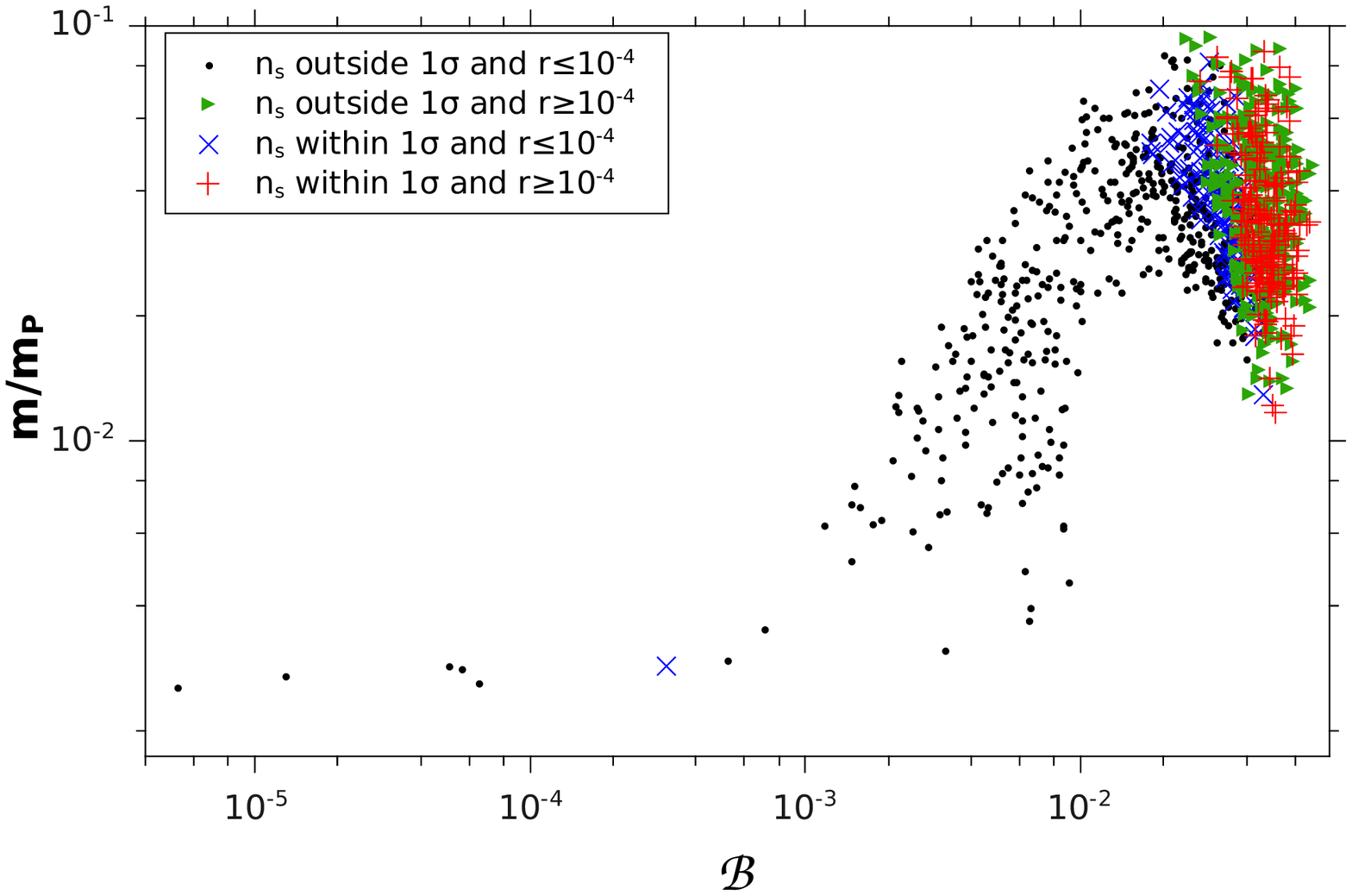} \\
		(c) & (d) \\
		\includegraphics[width=.5\columnwidth]{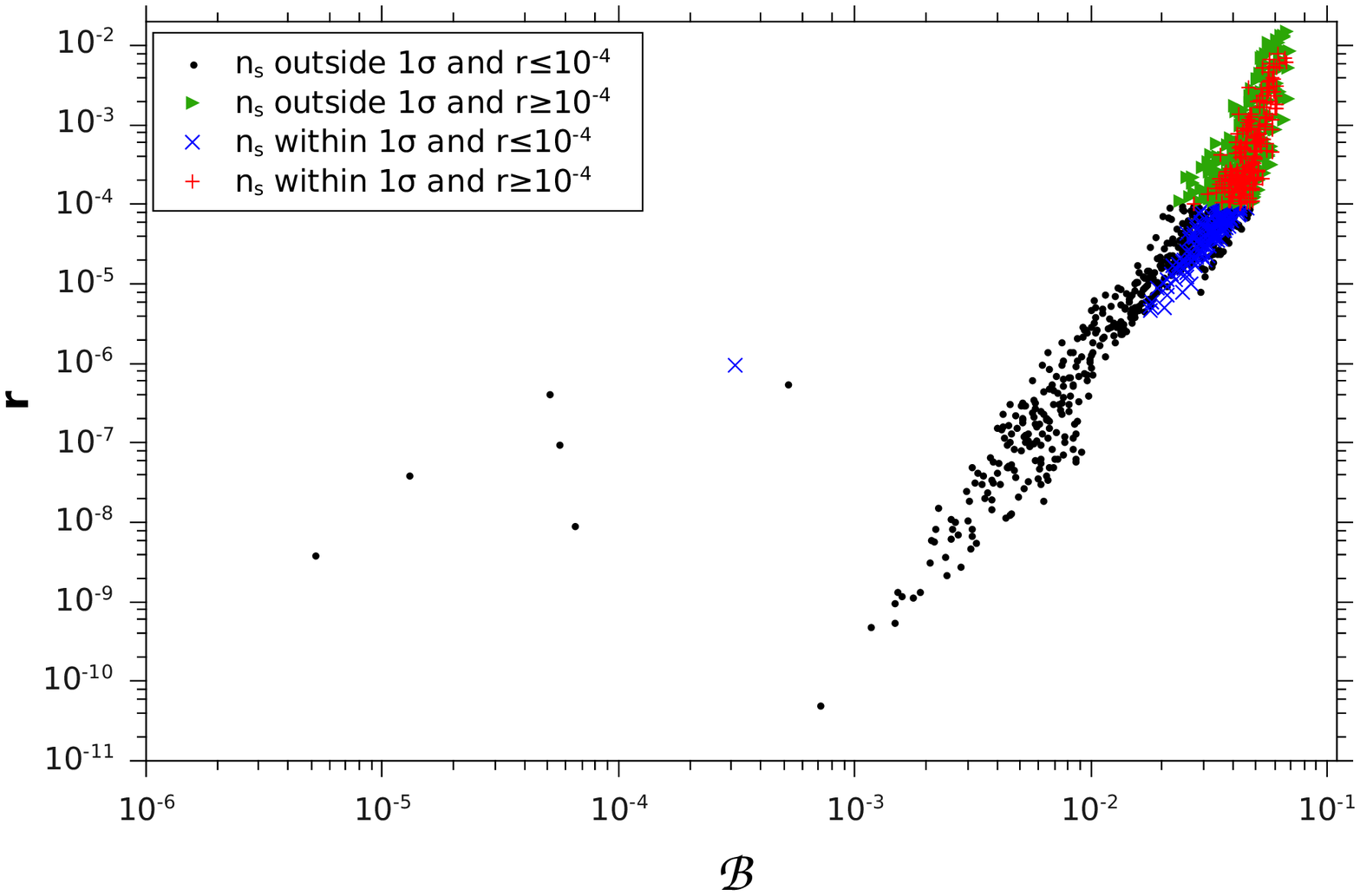} & \includegraphics[width=.5\columnwidth]{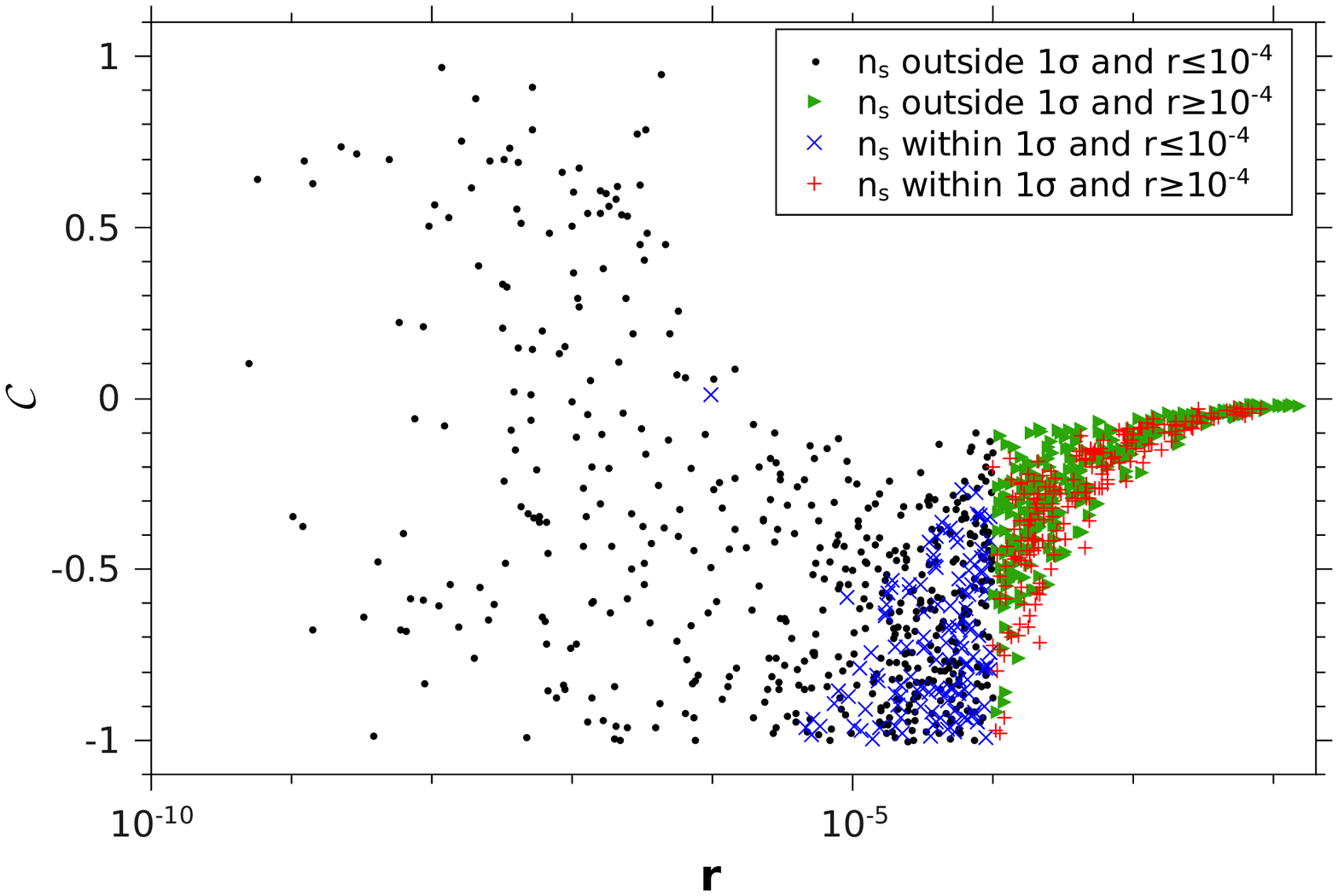} \\
		(e) & (f)
	\end{tabular}
	\caption{Results of our numerical calculations for the shifted hybrid inflation model with a non-minimal K\"ahler potential.  The color- and symbol-coding is intended to highlight various regions of interest, as laid out in the legend of each panel.}
	\label{NMKresults}
\end{figure}

We follow the same calculational techniques as those laid out in Refs.~\cite{Shafi:2010jr, Rehman:2010wm}.  We generate random values of the new parameters $\mathcal{A}, \mathcal{B}, \mathcal{C}$ within the ranges specified by Eqs.~(\ref{Amax})--(\ref{Cmin}), with the additional constraint that $|\mathcal{B}|, |\mathcal{C}| < \mathcal{A}$.  Given the expectations described above, we specialize to positive values of $\mathcal{B}$,\footnote{From preliminary analyses, we have seen empirically that successful inflation essentially occurs only for $\mathcal{B} > 0$, even when we have allowed for both signs.} and randomly generate $\log \mathcal{B}$ to better examine variation over multiple orders of magnitude.  For $\mathcal{A}$ and $\mathcal{C}$, we generate random values on a linear scale.  The ranges used to generate values for parameters in the potential are listed in Table~\ref{rangetable}, and are very similar to those used in the standard hybrid case.  Recall also that we fix $M_* = m_P$ in these calculations; since we expect the largest values of $r$ to occur near the Planck scale, we make this choice of cutoff in order to better probe how large $r$ may be in these models.

The results of our numerical calculations for the case of shifted inflation with a non-minimal K\"ahler potential are displayed in Fig.~\ref{NMKresults}.  These plots exhibit many similarities to the standard hybrid inflation results, but also some (largely quantitative) differences.  Panel~(a) of Fig.~\ref{NMKresults} shows the region in the $(n_s, r)$ plane where successful inflation may occur in this model.  We can see that, within the allowed range, these observables are essentially uncorrelated.  This panel also serves to define our color- and symbol-coding.  We have highlighted two regions of interest (as well as the overlap between these regions), namely points with large $r$-values ($\gtrsim 10^{-4}$) and points within a 1$\sigma$ range of the $n_s$ central value as given by WMAP7.\footnote{The 1$\sigma$ range referred to here has been extracted from the WMAP7 bounding curves in the $(n_s, r)$ plane.  In contrast, the values in Table~\ref{rangetable} used for the initial cuts were taken from the `best result' error bars quoted in the WMAP7 Cosmological Interpretation paper~\cite{Komatsu:2010fb}.  See also footnote~1 of Ref.~\cite{Shafi:2010jr}.}

As can be seen in various panels of Fig.~\ref{NMKresults}, we obtain values of the tensor-to-scalar ratio up to $r \sim 0.02$.  Values of $r$ on this order of magnitude are expected to be measurable by Planck, in contrast to those values exhibited by the minimal model examined in Section~\ref{sec_min}.  In this way, if Planck collects data with $r$ near the edge of its sensitivity ranges, the shifted hybrid inflation model may be made viable by the inclusion of higher order terms in the K\"ahler potential.

The prediction here for the largest value of $r$ is somewhat smaller than it was in the standard hybrid case, as given in Refs.~\cite{Shafi:2010jr, Rehman:2010wm}.  Recall that the largest $r$-values occur for field amplitudes closest to the Planck scale ($f \simeq 1$).  In the present case, we have taken this to mean $\sigma \simeq m_P$, but in the standard hybrid scenario it is more convenient to discuss the inflationary dynamics in terms of $|s| = \sigma/\sqrt{2}$.  Thus the previous treatments used $|s| \simeq m_P$ to define the Planck limit, which corresponds to allowing $\sigma$ values up to a factor $\sqrt{2}$ larger than those used here.

Panel~(b) shows, to some extent, the limiting behavior on $r$ due to the radiative correction contribution, as described above.  However, the expected decrease of $r$ at large values of $\kappa$ is not seen, due to the absence of successful points at sufficiently large $\kappa$.  As seen in Table~\ref{rangetable}, we have allowed for $\kappa$ to be somewhat larger than 1, yet the numerical results yield only $\kappa \lesssim 0.03$.  This can be understood by considering the relation
\begin{equation}
\kappa = \frac{\beta}{\xi} \left( \frac{M}{m_P} \right)^2 .
\label{kbeta}
\end{equation}
Naively, this seems to suggest that $\kappa$ increases with decreasing $\xi$.  However, some care must be taken here.  If we fix the value of $M$, a decrease in $\xi$ will result in an increase in $m$ (up to a maximum of $m = M$ for $\xi = 1/8$).  For successful inflation, the vacuum potential $V_0 \simeq \kappa^2 m^4$ is roughly constant near the GUT scale, so an increase in $m$ must in turn lead to a decrease in $\kappa$.  Thus the largest values of $\kappa$ should occur for the {\it largest} values of $\xi$.  (Indeed, this result can be shown more rigorously by eliminating $M$ in favor of $V_0$ in Eq.~(\ref{kbeta}), and noting that $V_0$ can no longer be treated as constant if $\xi$ becomes too close to $1/4$.)  Then, using $\beta \lesssim 1$ and $M/m_P \lesssim 0.1$, we obtain the upper limit of $\kappa \lesssim 0.04$.  Comparing to Fig.~\ref{NMKresults}, we see that this is essentially the limiting value that is exhibited in the results.

Since $m$ varies with $\kappa$ (roughly) as a power law, an upper bound on $\kappa$ also leads to a lower bound on $m$.  The bounding value obtained from $V_0 \sim$~constant is about $m/m_P \gtrsim 5 \times 10^{-3}$.  Panel~(c) of Fig.~\ref{NMKresults} shows that this value essentially holds.  It turns out that these limits on $\kappa$ and $m$ are largely responsible for the differences in the results of the present shifted hybrid model as compared to those of the standard hybrid scenario.  For example, Panel~(d) depicts the behavior of $m$ with respect to the quadratic coefficient $\mathcal{B}$.  We anticipate that the model should reduce to its global SUSY version in the limit where the SUGRA couplings vanish, which coincides with the limit as $\mathcal{B} \rightarrow 0$.  From Ref.~\cite{Rehman:2010wm}, this behavior was exhibited quite clearly in the standard hybrid model, where the mass parameter $M$ tended toward the global SUSY prediction of $\sim$~GUT scale in this limit.  We expect that $m$ should also tend toward a constant value around the same scale; while we do see he beginnings of such behavior, the point density is low due to the limiting behavior described above, and we cannot draw a firm conclusion in the present model.

In Panel~(e), we display the behavior of $r$ with respect to $\mathcal{B}$.  On a log-log scale, this variation is strikingly linear.  We recall that the previous treatments of the standard hybrid case revealed increasing behavior between $r$ and the quadratic coefficient as well, at which time this was attributed to the need for increased dominance over the quartic term as $f$ increases.  In that case, though, the plot increased linearly only at the largest values of $r$; in the shifted model, we see that this dependence is much sharper.

Other differences versus the standard hybrid scenario may be attributed to small differences in the definitions of such parameters as $x$ and the coefficients of various terms in the potential, which we have already noted.  There is also a subtle change in the way the quartic coefficient was generated; in the standard case, the fundamental couplings $\kappa_S$ and $\kappa_{SS}$ were generated (on a log scale) and then served to specify the quartic coefficient, whereas in the shifted model we have directly generated $\mathcal{C}$ (on a linear scale).  In addition, the allowed range here is somewhat more limited by the constraint $|\mathcal{C}| \lesssim \mathcal{A} \sim 1$.  The variation of this quartic coefficient with respect to $r$ is displayed in Panel~(f) of Fig.~\ref{NMKresults}.

It is worthwhile to comment on possible issues and the difficulties that were experienced in the standard hybrid inflation model.  Given the high degree of similarity between the inflationary potentials of the standard and shifted hybrid scenarios, many of the issues exhibited before will persist here and may be handled in a similar way as in the standard case.  However, there is one aspect in which the shifted model has an advantage over the standard hybrid model.  As we have already noted, the motivation for employing shifted hybrid inflation is primarily to deal with problematic topological defects.  In standard SUSY hybrid inflation, any topological defect production due to the breaking of the gauge group $G$ occurs at the end of inflation, when the system undergoes a waterfall transition.  Depending on the gauge group, these defects may be cosmic strings (e.g. from $G = \text{U(1)}$), whose density must be sufficiently suppressed~\cite{Battye:2010xz} to agree with observations, or monopoles (from models such as $G = \text{SU(4)}\times\text{SU(2)}\times\text{SU(2)}$ and $G = \text{SU(5)}$) which must essentially be inflated away.  By choosing the shifted track for the entire duration of observable inflation, we have ensured that these objects experience the required suppression in density.  Indeed, in this article, we have shown that this issue may be resolved without a substantial change to the predictions of observable parameters that are being measured by current satellite observatories.


\section{Summary}
\label{sec_sum}

We have provided an extensive update to models of shifted supersymmetric hybrid inflation, where we have included contributions from the supergravity and soft SUSY-breaking sectors in addition to the usual radiative corrections.  If one of the relevant soft terms is negative, a red-tilted spectrum with $n_s$ spanning the WMAP7 2$\sigma$ range may be obtained with the use of the canonical K\"ahler potential.  If higher order corrections are included in the K\"ahler potential, the possibility of a red-tilted spectrum persists, and the tensor modes can attain values which are substantially larger.  Indeed, the tensor-to-scalar ratio in this version of the model may reach values as large as $r \simeq 0.02$, which is potentially observable by the Planck Satellite.  In contrast to the standard hybrid scenario, the shifted model ensures that any topological defects produced in the breaking of the gauge symmetry are inflated away.  It should be interesting to extend our discussion to other inflationary scenarios, such as smooth hybrid inflation~\cite{Lazarides:1995vr, *Lazarides:1996rk, *Jeannerot:2001qu, Senoguz:2004vu, urRehman:2006hu} and warm inflation~\cite{*[{}]  [{ and references therein.}] BasteroGil:2009ec} models.

\begin{acknowledgments}

This work is supported in part by the DOE under grant No.~DE-FG02-91ER40626, by the University of Delaware (M.R.), and by NASA and the Delaware Space Grant Consortium under grant No.\ NNG05GO92H (J.W.).

\end{acknowledgments}


\appendix*
\section{Functional Forms of $\mathcal{A}, \mathcal{B}, \mathcal{C}$}
\label{app}

Before writing down the expressions for $\mathcal{A}, \mathcal{B}, \mathcal{C}$, it is convenient to define some auxiliary functions of the couplings in the K\"ahler potential:
\begin{eqnarray*}
c_0 &=& 1 - \frac{1}{2} ( \kappa_{S\Phi} + \kappa_{S\overline{\Phi}} ) , \\
c_1 &=& 1 + \frac{1}{8} \left[ 4 \kappa_{S\Phi}^2 - \kappa_{S\Phi\Phi} - 4 \kappa_{S\Phi\overline{\Phi}} + 8 \kappa_{S\Phi} (-1 + \kappa_{S\overline{\Phi}}) \right. \\
  && \quad\quad \left. + 4 (-2 + \kappa_{S\overline{\Phi}}) \kappa_{S\overline{\Phi}} - \kappa_{S\overline{\Phi}\overline{\Phi}} + \kappa_{\Phi} + 4 \kappa_{\Phi\overline{\Phi}} + \kappa_{\overline{\Phi}} \right] , \\
c_2 &=& 1 + \frac{1}{2} \left[ - \kappa_{SS\Phi} - \kappa_{SS\overline{\Phi}} + (-2 + \kappa_{S\Phi}) \kappa_{S\Phi} + (-2 + \kappa_{S\overline{\Phi}}) \kappa_{S\overline{\Phi}} \right. \\
  && \quad\quad \left. + 2 \kappa_S (-1 + \kappa_{S\Phi} + \kappa_{S\overline{\Phi}}) \right] .
\end{eqnarray*}
Notice that each of these reduces to 1 in the case of minimal K\"ahler.  We may also write the function
\begin{equation*}
\gamma_S = 1 - \frac{7}{2}\kappa_S + 2\kappa_S^2 - 3\kappa_{SS} ,
\end{equation*}
which is the same parameter that appears in the quartic coefficient of the standard hybrid inflation model, and also reduces to 1 for minimal K\"ahler.  In the case of non-minimal K\"ahler with couplings $-1 \lesssim \kappa_i \lesssim 1$, we obtain the ranges
\begin{eqnarray}
0 \lesssim &c_0& \lesssim 2 , \nonumber \\
-1 \lesssim &c_1& \lesssim \frac{13}{2} , \label{cranges} \\
-2 \lesssim &c_2& \lesssim 8 , \nonumber \\
-\frac{113}{32} \lesssim &\gamma_S& \lesssim \frac{19}{2} . \nonumber
\end{eqnarray}
This range for the $\kappa_i$ couplings is somewhat more restrictive than needed; for perturbativity, we must only have $|\kappa_i| \lesssim O(1)$, and there is some ambiguity involved with how the combinatoric factors are written down in the K\"ahler potential.  Looking slightly ahead, it will be most convenient if we are able to treat $c_0, c_1, c_2$ and $\gamma_S$ as independently varying parameters, so that the quantities $\mathcal{A}, \mathcal{B}, \mathcal{C}$ may be varied independently.  From the definitions of these $c_i$'s above, one can readily verify that this is not quite the case within the specified ranges; there exist some interdependencies via the underlying couplings, which lead to some regions of the $(c_0, c_1, c_2)$ being impossible to access.  However, relaxing to somewhat larger values of the $|\kappa_i|$'s (but still order 1) has the effect of $c_0, c_1, c_2$ becoming essentially independent within their ranges in Eqs.~(\ref{cranges}).  Based on these arguments, and for concreteness and simplicity, we will take these ranges for $c_0, c_1, c_2$, and assume that they may vary independently within these ranges.

Now, we define $\mathcal{A}, \mathcal{B}$, and $\mathcal{C}$ as the coefficients of the terms constant, quadratic, and quartic (respectively) in $|s|/m_P$ in the normalized potential $V/\kappa^2 m^4$.  These quantities appear in the form
\begin{eqnarray*}
\mathcal{A} &=& 1 + 2 c_0 \phi_P^2 + 2 c_1 \phi_P^4 , \\
\mathcal{B} &=& -\kappa_S + 2 c_2 \phi_P^2 , \\
\mathcal{C} &=& \frac{\gamma_S}{2} .
\end{eqnarray*}
Again, we take these parameters as independent of one another (although this assumption only approximately holds, to within factors of order unity), which makes the numerical calculations more tractable.  Using these expressions with the ranges in Eqs.~(\ref{cranges}), the extremal functions in Eqs.~(\ref{Amax})--(\ref{Cmin}) can easily be verified.

\bibliographystyle{apsrev4-1}
\bibliography{shiftedref}

\begin{thebibliography}{30}%
\makeatletter
\providecommand \@ifxundefined [1]{%
 \@ifx{#1\undefined}
}%
\providecommand \@ifnum [1]{%
 \ifnum #1\expandafter \@firstoftwo
 \else \expandafter \@secondoftwo
 \fi
}%
\providecommand \@ifx [1]{%
 \ifx #1\expandafter \@firstoftwo
 \else \expandafter \@secondoftwo
 \fi
}%
\providecommand \natexlab [1]{#1}%
\providecommand \enquote  [1]{``#1''}%
\providecommand \bibnamefont  [1]{#1}%
\providecommand \bibfnamefont [1]{#1}%
\providecommand \citenamefont [1]{#1}%
\providecommand \href@noop [0]{\@secondoftwo}%
\providecommand \href [0]{\begingroup \@sanitize@url \@href}%
\providecommand \@href[1]{\@@startlink{#1}\@@href}%
\providecommand \@@href[1]{\endgroup#1\@@endlink}%
\providecommand \@sanitize@url [0]{\catcode `\\12\catcode `\$12\catcode
  `\&12\catcode `\#12\catcode `\^12\catcode `\_12\catcode `\%12\relax}%
\providecommand \@@startlink[1]{}%
\providecommand \@@endlink[0]{}%
\providecommand \url  [0]{\begingroup\@sanitize@url \@url }%
\providecommand \@url [1]{\endgroup\@href {#1}{\urlprefix }}%
\providecommand \urlprefix  [0]{URL }%
\providecommand \Eprint [0]{\href }%
\providecommand \doibase [0]{http://dx.doi.org/}%
\providecommand \selectlanguage [0]{\@gobble}%
\providecommand \bibinfo  [0]{\@secondoftwo}%
\providecommand \bibfield  [0]{\@secondoftwo}%
\providecommand \translation [1]{[#1]}%
\providecommand \BibitemOpen [0]{}%
\providecommand \bibitemStop [0]{}%
\providecommand \bibitemNoStop [0]{.\EOS\space}%
\providecommand \EOS [0]{\spacefactor3000\relax}%
\providecommand \BibitemShut  [1]{\csname bibitem#1\endcsname}%
\let\auto@bib@innerbib\@empty
\bibitem [{\citenamefont {Dvali}\ \emph {et~al.}(1994)\citenamefont {Dvali},
  \citenamefont {Shafi},\ and\ \citenamefont {Schaefer}}]{Dvali:1994ms}%
  \BibitemOpen
  \bibfield  {author} {\bibinfo {author} {\bibfnamefont {G.~R.}\ \bibnamefont
  {Dvali}}, \bibinfo {author} {\bibfnamefont {Q.}~\bibnamefont {Shafi}}, \ and\
  \bibinfo {author} {\bibfnamefont {R.~K.}\ \bibnamefont {Schaefer}},\ }\href
  {\doibase 10.1103/PhysRevLett.73.1886} {\bibfield  {journal} {\bibinfo
  {journal} {Phys. Rev. Lett.}\ }\textbf {\bibinfo {volume} {73}},\ \bibinfo
  {pages} {1886} (\bibinfo {year} {1994})},\ \Eprint
  {http://arxiv.org/abs/hep-ph/9406319} {arXiv:hep-ph/9406319} \BibitemShut
  {NoStop}%
\bibitem [{\citenamefont {Copeland}\ \emph {et~al.}(1994)\citenamefont
  {Copeland}, \citenamefont {Liddle}, \citenamefont {Lyth}, \citenamefont
  {Stewart},\ and\ \citenamefont {Wands}}]{Copeland:1994vg}%
  \BibitemOpen
  \bibfield  {author} {\bibinfo {author} {\bibfnamefont {E.~J.}\ \bibnamefont
  {Copeland}}, \bibinfo {author} {\bibfnamefont {A.~R.}\ \bibnamefont
  {Liddle}}, \bibinfo {author} {\bibfnamefont {D.~H.}\ \bibnamefont {Lyth}},
  \bibinfo {author} {\bibfnamefont {E.~D.}\ \bibnamefont {Stewart}}, \ and\
  \bibinfo {author} {\bibfnamefont {D.}~\bibnamefont {Wands}},\ }\href
  {\doibase 10.1103/PhysRevD.49.6410} {\bibfield  {journal} {\bibinfo
  {journal} {Phys. Rev.}\ }\textbf {\bibinfo {volume} {D49}},\ \bibinfo {pages}
  {6410} (\bibinfo {year} {1994})},\ \Eprint
  {http://arxiv.org/abs/astro-ph/9401011} {arXiv:astro-ph/9401011} \BibitemShut
  {NoStop}%
\bibitem [{\citenamefont {Senoguz}\ and\ \citenamefont
  {Shafi}(2003)}]{Senoguz:2003zw}%
  \BibitemOpen
  \bibfield  {author} {\bibinfo {author} {\bibfnamefont {V.~N.}\ \bibnamefont
  {Senoguz}}\ and\ \bibinfo {author} {\bibfnamefont {Q.}~\bibnamefont
  {Shafi}},\ }\href {\doibase 10.1016/j.physletb.2003.06.030} {\bibfield
  {journal} {\bibinfo  {journal} {Phys. Lett.}\ }\textbf {\bibinfo {volume}
  {B567}},\ \bibinfo {pages} {79} (\bibinfo {year} {2003})},\ \Eprint
  {http://arxiv.org/abs/hep-ph/0305089} {arXiv:hep-ph/0305089} \BibitemShut
  {NoStop}%
\bibitem [{\citenamefont {Panagiotakopoulos}(1997)}]{Panagiotakopoulos:1997qd}%
  \BibitemOpen
  \bibfield  {author} {\bibinfo {author} {\bibfnamefont {C.}~\bibnamefont
  {Panagiotakopoulos}},\ }\href {\doibase 10.1103/PhysRevD.55.R7335} {\bibfield
   {journal} {\bibinfo  {journal} {Phys. Rev.}\ }\textbf {\bibinfo {volume}
  {D55}},\ \bibinfo {pages} {7335} (\bibinfo {year} {1997})},\ \Eprint
  {http://arxiv.org/abs/hep-ph/9702433} {arXiv:hep-ph/9702433} \BibitemShut
  {NoStop}%
\bibitem [{\citenamefont {Linde}\ and\ \citenamefont
  {Riotto}(1997)}]{Linde:1997sj}%
  \BibitemOpen
  \bibfield  {author} {\bibinfo {author} {\bibfnamefont {A.~D.}\ \bibnamefont
  {Linde}}\ and\ \bibinfo {author} {\bibfnamefont {A.}~\bibnamefont {Riotto}},\
  }\href {\doibase 10.1103/PhysRevD.56.R1841} {\bibfield  {journal} {\bibinfo
  {journal} {Phys. Rev.}\ }\textbf {\bibinfo {volume} {D56}},\ \bibinfo {pages}
  {1841} (\bibinfo {year} {1997})},\ \Eprint
  {http://arxiv.org/abs/hep-ph/9703209} {arXiv:hep-ph/9703209} \BibitemShut
  {NoStop}%
\bibitem [{\citenamefont {Buchmuller}\ \emph {et~al.}(2000)\citenamefont
  {Buchmuller}, \citenamefont {Covi},\ and\ \citenamefont
  {Delepine}}]{Buchmuller:2000zm}%
  \BibitemOpen
  \bibfield  {author} {\bibinfo {author} {\bibfnamefont {W.}~\bibnamefont
  {Buchmuller}}, \bibinfo {author} {\bibfnamefont {L.}~\bibnamefont {Covi}}, \
  and\ \bibinfo {author} {\bibfnamefont {D.}~\bibnamefont {Delepine}},\ }\href
  {\doibase 10.1016/S0370-2693(00)01005-4} {\bibfield  {journal} {\bibinfo
  {journal} {Phys. Lett.}\ }\textbf {\bibinfo {volume} {B491}},\ \bibinfo
  {pages} {183} (\bibinfo {year} {2000})},\ \Eprint
  {http://arxiv.org/abs/hep-ph/0006168} {arXiv:hep-ph/0006168} \BibitemShut
  {NoStop}%
\bibitem [{\citenamefont {Kawasaki}\ \emph {et~al.}(2003)\citenamefont
  {Kawasaki}, \citenamefont {Yamaguchi},\ and\ \citenamefont
  {Yokoyama}}]{Kawasaki:2003zv}%
  \BibitemOpen
  \bibfield  {author} {\bibinfo {author} {\bibfnamefont {M.}~\bibnamefont
  {Kawasaki}}, \bibinfo {author} {\bibfnamefont {M.}~\bibnamefont {Yamaguchi}},
  \ and\ \bibinfo {author} {\bibfnamefont {J.}~\bibnamefont {Yokoyama}},\
  }\href {\doibase 10.1103/PhysRevD.68.023508} {\bibfield  {journal} {\bibinfo
  {journal} {Phys. Rev.}\ }\textbf {\bibinfo {volume} {D68}},\ \bibinfo {pages}
  {023508} (\bibinfo {year} {2003})},\ \Eprint
  {http://arxiv.org/abs/hep-ph/0304161} {arXiv:hep-ph/0304161} \BibitemShut
  {NoStop}%
\bibitem [{\citenamefont {Komatsu}\ \emph {et~al.}(2011)\citenamefont {Komatsu}
  \emph {et~al.}}]{Komatsu:2010fb}%
  \BibitemOpen
  \bibfield  {author} {\bibinfo {author} {\bibfnamefont {E.}~\bibnamefont
  {Komatsu}} \emph {et~al.} (\bibinfo {collaboration} {WMAP}),\ }\href
  {\doibase 10.1088/0067-0049/192/2/18} {\bibfield  {journal} {\bibinfo
  {journal} {Astrophys. J. Suppl.}\ }\textbf {\bibinfo {volume} {192}},\
  \bibinfo {pages} {18} (\bibinfo {year} {2011})},\ \Eprint
  {http://arxiv.org/abs/1001.4538} {arXiv:1001.4538 [astro-ph.CO]} \BibitemShut
  {NoStop}%
\bibitem [{\citenamefont {Senoguz}\ and\ \citenamefont
  {Shafi}(2005)}]{Senoguz:2004vu}%
  \BibitemOpen
  \bibfield  {author} {\bibinfo {author} {\bibfnamefont {V.~N.}\ \bibnamefont
  {Senoguz}}\ and\ \bibinfo {author} {\bibfnamefont {Q.}~\bibnamefont
  {Shafi}},\ }\href {\doibase 10.1103/PhysRevD.71.043514} {\bibfield  {journal}
  {\bibinfo  {journal} {Phys. Rev.}\ }\textbf {\bibinfo {volume} {D71}},\
  \bibinfo {pages} {043514} (\bibinfo {year} {2005})},\ \Eprint
  {http://arxiv.org/abs/hep-ph/0412102} {arXiv:hep-ph/0412102} \BibitemShut
  {NoStop}%
\bibitem [{\citenamefont {Jeannerot}\ and\ \citenamefont
  {Postma}(2005)}]{Jeannerot:2005mc}%
  \BibitemOpen
  \bibfield  {author} {\bibinfo {author} {\bibfnamefont {R.}~\bibnamefont
  {Jeannerot}}\ and\ \bibinfo {author} {\bibfnamefont {M.}~\bibnamefont
  {Postma}},\ }\href@noop {} {\bibfield  {journal} {\bibinfo  {journal} {JHEP}\
  }\textbf {\bibinfo {volume} {05}},\ \bibinfo {pages} {071} (\bibinfo {year}
  {2005})},\ \Eprint {http://arxiv.org/abs/hep-ph/0503146}
  {arXiv:hep-ph/0503146} \BibitemShut {NoStop}%
\bibitem [{\citenamefont {Rehman}\ \emph
  {et~al.}(2010{\natexlab{a}})\citenamefont {Rehman}, \citenamefont {Shafi},\
  and\ \citenamefont {Wickman}}]{Rehman:2009nq}%
  \BibitemOpen
  \bibfield  {author} {\bibinfo {author} {\bibfnamefont {M.~U.}\ \bibnamefont
  {Rehman}}, \bibinfo {author} {\bibfnamefont {Q.}~\bibnamefont {Shafi}}, \
  and\ \bibinfo {author} {\bibfnamefont {J.~R.}\ \bibnamefont {Wickman}},\
  }\href {\doibase 10.1016/j.physletb.2009.12.010} {\bibfield  {journal}
  {\bibinfo  {journal} {Phys.Lett.}\ }\textbf {\bibinfo {volume} {B683}},\
  \bibinfo {pages} {191} (\bibinfo {year} {2010}{\natexlab{a}})},\ \Eprint
  {http://arxiv.org/abs/0908.3896} {arXiv:0908.3896 [hep-ph]} \BibitemShut
  {NoStop}%
\bibitem [{\citenamefont {Rehman}\ \emph
  {et~al.}(2010{\natexlab{b}})\citenamefont {Rehman}, \citenamefont {Shafi},\
  and\ \citenamefont {Wickman}}]{Rehman:2009yj}%
  \BibitemOpen
  \bibfield  {author} {\bibinfo {author} {\bibfnamefont {M.~U.}\ \bibnamefont
  {Rehman}}, \bibinfo {author} {\bibfnamefont {Q.}~\bibnamefont {Shafi}}, \
  and\ \bibinfo {author} {\bibfnamefont {J.~R.}\ \bibnamefont {Wickman}},\
  }\href {\doibase 10.1016/j.physletb.2010.03.072} {\bibfield  {journal}
  {\bibinfo  {journal} {Phys.Lett.}\ }\textbf {\bibinfo {volume} {B688}},\
  \bibinfo {pages} {75} (\bibinfo {year} {2010}{\natexlab{b}})},\ \Eprint
  {http://arxiv.org/abs/0912.4737} {arXiv:0912.4737 [hep-ph]} \BibitemShut
  {NoStop}%
\bibitem [{\citenamefont {Pallis}(2007)}]{Pallis:2007du}%
  \BibitemOpen
  \bibfield  {author} {\bibinfo {author} {\bibfnamefont {C.}~\bibnamefont
  {Pallis}},\ }\href@noop {} {\  (\bibinfo {year} {2007})},\ \Eprint
  {http://arxiv.org/abs/0710.3074} {arXiv:0710.3074 [hep-ph]} \BibitemShut
  {NoStop}%
\bibitem [{\citenamefont {Bastero-Gil}\ \emph {et~al.}(2007)\citenamefont
  {Bastero-Gil}, \citenamefont {King},\ and\ \citenamefont
  {Shafi}}]{BasteroGil:2006cm}%
  \BibitemOpen
  \bibfield  {author} {\bibinfo {author} {\bibfnamefont {M.}~\bibnamefont
  {Bastero-Gil}}, \bibinfo {author} {\bibfnamefont {S.~F.}\ \bibnamefont
  {King}}, \ and\ \bibinfo {author} {\bibfnamefont {Q.}~\bibnamefont {Shafi}},\
  }\href {\doibase 10.1016/j.physletb.2006.06.085} {\bibfield  {journal}
  {\bibinfo  {journal} {Phys. Lett.}\ }\textbf {\bibinfo {volume} {B651}},\
  \bibinfo {pages} {345} (\bibinfo {year} {2007})},\ \Eprint
  {http://arxiv.org/abs/hep-ph/0604198} {arXiv:hep-ph/0604198} \BibitemShut
  {NoStop}%
\bibitem [{\citenamefont {ur~Rehman}\ \emph {et~al.}(2007)\citenamefont
  {ur~Rehman}, \citenamefont {Senoguz},\ and\ \citenamefont
  {Shafi}}]{urRehman:2006hu}%
  \BibitemOpen
  \bibfield  {author} {\bibinfo {author} {\bibfnamefont {M.}~\bibnamefont
  {ur~Rehman}}, \bibinfo {author} {\bibfnamefont {V.~N.}\ \bibnamefont
  {Senoguz}}, \ and\ \bibinfo {author} {\bibfnamefont {Q.}~\bibnamefont
  {Shafi}},\ }\href {\doibase 10.1103/PhysRevD.75.043522} {\bibfield  {journal}
  {\bibinfo  {journal} {Phys. Rev.}\ }\textbf {\bibinfo {volume} {D75}},\
  \bibinfo {pages} {043522} (\bibinfo {year} {2007})},\ \Eprint
  {http://arxiv.org/abs/hep-ph/0612023} {arXiv:hep-ph/0612023} \BibitemShut
  {NoStop}%
\bibitem [{\citenamefont {Shafi}\ and\ \citenamefont
  {Wickman}(2011)}]{Shafi:2010jr}%
  \BibitemOpen
  \bibfield  {author} {\bibinfo {author} {\bibfnamefont {Q.}~\bibnamefont
  {Shafi}}\ and\ \bibinfo {author} {\bibfnamefont {J.~R.}\ \bibnamefont
  {Wickman}},\ }\href {\doibase 10.1016/j.physletb.2011.01.002} {\bibfield
  {journal} {\bibinfo  {journal} {Phys.Lett.}\ }\textbf {\bibinfo {volume}
  {B696}},\ \bibinfo {pages} {438} (\bibinfo {year} {2011})},\ \Eprint
  {http://arxiv.org/abs/1009.5340} {arXiv:1009.5340 [hep-ph]} \BibitemShut
  {NoStop}%
\bibitem [{\citenamefont {Rehman}\ \emph {et~al.}(2011)\citenamefont {Rehman},
  \citenamefont {Shafi},\ and\ \citenamefont {Wickman}}]{Rehman:2010wm}%
  \BibitemOpen
  \bibfield  {author} {\bibinfo {author} {\bibfnamefont {M.~U.}\ \bibnamefont
  {Rehman}}, \bibinfo {author} {\bibfnamefont {Q.}~\bibnamefont {Shafi}}, \
  and\ \bibinfo {author} {\bibfnamefont {J.~R.}\ \bibnamefont {Wickman}},\
  }\href {\doibase 10.1103/PhysRevD.83.067304} {\bibfield  {journal} {\bibinfo
  {journal} {Phys. Rev.}\ }\textbf {\bibinfo {volume} {D83}},\ \bibinfo {pages}
  {067304} (\bibinfo {year} {2011})},\ \Eprint {http://arxiv.org/abs/1012.0309}
  {arXiv:1012.0309 [astro-ph.CO]} \BibitemShut {NoStop}%
\bibitem [{\citenamefont {Jeannerot}\ \emph {et~al.}(2000)\citenamefont
  {Jeannerot}, \citenamefont {Khalil}, \citenamefont {Lazarides},\ and\
  \citenamefont {Shafi}}]{Jeannerot:2000sv}%
  \BibitemOpen
  \bibfield  {author} {\bibinfo {author} {\bibfnamefont {R.}~\bibnamefont
  {Jeannerot}}, \bibinfo {author} {\bibfnamefont {S.}~\bibnamefont {Khalil}},
  \bibinfo {author} {\bibfnamefont {G.}~\bibnamefont {Lazarides}}, \ and\
  \bibinfo {author} {\bibfnamefont {Q.}~\bibnamefont {Shafi}},\ }\href@noop {}
  {\bibfield  {journal} {\bibinfo  {journal} {JHEP}\ }\textbf {\bibinfo
  {volume} {10}},\ \bibinfo {pages} {012} (\bibinfo {year} {2000})},\ \Eprint
  {http://arxiv.org/abs/hep-ph/0002151} {arXiv:hep-ph/0002151} \BibitemShut
  {NoStop}%
\bibitem [{\citenamefont {Antusch}\ \emph {et~al.}(2010)\citenamefont {Antusch}
  \emph {et~al.}}]{Antusch:2010va}%
  \BibitemOpen
  \bibfield  {author} {\bibinfo {author} {\bibfnamefont {S.}~\bibnamefont
  {Antusch}} \emph {et~al.},\ }\href {\doibase 10.1007/JHEP08(2010)100}
  {\bibfield  {journal} {\bibinfo  {journal} {JHEP}\ }\textbf {\bibinfo
  {volume} {08}},\ \bibinfo {pages} {100} (\bibinfo {year} {2010})},\ \Eprint
  {http://arxiv.org/abs/1003.3233} {arXiv:1003.3233 [hep-ph]} \BibitemShut
  {NoStop}%
\bibitem [{\citenamefont {Senoguz}\ and\ \citenamefont
  {Shafi}(2004)}]{Senoguz:2004ky}%
  \BibitemOpen
  \bibfield  {author} {\bibinfo {author} {\bibfnamefont {V.~N.}\ \bibnamefont
  {Senoguz}}\ and\ \bibinfo {author} {\bibfnamefont {Q.}~\bibnamefont
  {Shafi}},\ }\href {\doibase 10.1016/j.physletb.2004.05.077} {\bibfield
  {journal} {\bibinfo  {journal} {Phys. Lett.}\ }\textbf {\bibinfo {volume}
  {B596}},\ \bibinfo {pages} {8} (\bibinfo {year} {2004})},\ \Eprint
  {http://arxiv.org/abs/hep-ph/0403294} {arXiv:hep-ph/0403294} \BibitemShut
  {NoStop}%
\bibitem [{\citenamefont {Allahverdi}\ \emph {et~al.}(2006)\citenamefont
  {Allahverdi}, \citenamefont {Enqvist}, \citenamefont {Garcia-Bellido},\ and\
  \citenamefont {Mazumdar}}]{Allahverdi:2006iq}%
  \BibitemOpen
  \bibfield  {author} {\bibinfo {author} {\bibfnamefont {R.}~\bibnamefont
  {Allahverdi}}, \bibinfo {author} {\bibfnamefont {K.}~\bibnamefont {Enqvist}},
  \bibinfo {author} {\bibfnamefont {J.}~\bibnamefont {Garcia-Bellido}}, \ and\
  \bibinfo {author} {\bibfnamefont {A.}~\bibnamefont {Mazumdar}},\ }\href
  {\doibase 10.1103/PhysRevLett.97.191304} {\bibfield  {journal} {\bibinfo
  {journal} {Phys. Rev. Lett.}\ }\textbf {\bibinfo {volume} {97}},\ \bibinfo
  {pages} {191304} (\bibinfo {year} {2006})},\ \Eprint
  {http://arxiv.org/abs/hep-ph/0605035} {arXiv:hep-ph/0605035} \BibitemShut
  {NoStop}%
\bibitem [{\citenamefont {Khalil}\ \emph {et~al.}(2011)\citenamefont {Khalil},
  \citenamefont {Rehman}, \citenamefont {Shafi},\ and\ \citenamefont
  {Zaakouk}}]{Khalil:2010cp}%
  \BibitemOpen
  \bibfield  {author} {\bibinfo {author} {\bibfnamefont {S.}~\bibnamefont
  {Khalil}}, \bibinfo {author} {\bibfnamefont {M.~U.}\ \bibnamefont {Rehman}},
  \bibinfo {author} {\bibfnamefont {Q.}~\bibnamefont {Shafi}}, \ and\ \bibinfo
  {author} {\bibfnamefont {E.~A.}\ \bibnamefont {Zaakouk}},\ }\href@noop {}
  {\bibfield  {journal} {\bibinfo  {journal} {Phys. Rev.}\ }\textbf {\bibinfo
  {volume} {D83}},\ \bibinfo {pages} {063522} (\bibinfo {year} {2011})},\
  \Eprint {http://arxiv.org/abs/1010.3657} {arXiv:1010.3657 [hep-ph]}
  \BibitemShut {NoStop}%
\bibitem [{\citenamefont {Coleman}\ and\ \citenamefont
  {Weinberg}(1973)}]{Coleman:1973jx}%
  \BibitemOpen
  \bibfield  {author} {\bibinfo {author} {\bibfnamefont {S.~R.}\ \bibnamefont
  {Coleman}}\ and\ \bibinfo {author} {\bibfnamefont {E.~J.}\ \bibnamefont
  {Weinberg}},\ }\href {\doibase 10.1103/PhysRevD.7.1888} {\bibfield  {journal}
  {\bibinfo  {journal} {Phys.Rev.}\ }\textbf {\bibinfo {volume} {D7}},\
  \bibinfo {pages} {1888} (\bibinfo {year} {1973})}\BibitemShut {NoStop}%
\bibitem [{\citenamefont {Boubekeur}\ and\ \citenamefont
  {Lyth}(2005)}]{Boubekeur:2005zm}%
  \BibitemOpen
  \bibfield  {author} {\bibinfo {author} {\bibfnamefont {L.}~\bibnamefont
  {Boubekeur}}\ and\ \bibinfo {author} {\bibfnamefont {D.~H.}\ \bibnamefont
  {Lyth}},\ }\href {\doibase 10.1088/1475-7516/2005/07/010} {\bibfield
  {journal} {\bibinfo  {journal} {JCAP}\ }\textbf {\bibinfo {volume} {0507}},\
  \bibinfo {pages} {010} (\bibinfo {year} {2005})},\ \Eprint
  {http://arxiv.org/abs/hep-ph/0502047} {arXiv:hep-ph/0502047} \BibitemShut
  {NoStop}%
\bibitem [{\citenamefont {Pallis}(2009)}]{Pallis:2009pq}%
  \BibitemOpen
  \bibfield  {author} {\bibinfo {author} {\bibfnamefont {C.}~\bibnamefont
  {Pallis}},\ }\href {\doibase 10.1088/1475-7516/2009/04/024} {\bibfield
  {journal} {\bibinfo  {journal} {JCAP}\ }\textbf {\bibinfo {volume} {0904}},\
  \bibinfo {pages} {024} (\bibinfo {year} {2009})},\ \Eprint
  {http://arxiv.org/abs/0902.0334} {arXiv:0902.0334 [hep-ph]} \BibitemShut
  {NoStop}%
\bibitem [{\citenamefont {Battye}\ and\ \citenamefont
  {Moss}(2010)}]{Battye:2010xz}%
  \BibitemOpen
  \bibfield  {author} {\bibinfo {author} {\bibfnamefont {R.}~\bibnamefont
  {Battye}}\ and\ \bibinfo {author} {\bibfnamefont {A.}~\bibnamefont {Moss}},\
  }\href {\doibase 10.1103/PhysRevD.82.023521} {\bibfield  {journal} {\bibinfo
  {journal} {Phys. Rev.}\ }\textbf {\bibinfo {volume} {D82}},\ \bibinfo {pages}
  {023521} (\bibinfo {year} {2010})},\ \Eprint {http://arxiv.org/abs/1005.0479}
  {arXiv:1005.0479 [astro-ph.CO]} \BibitemShut {NoStop}%
\bibitem [{\citenamefont {Lazarides}\ and\ \citenamefont
  {Panagiotakopoulos}(1995)}]{Lazarides:1995vr}%
  \BibitemOpen
  \bibfield  {author} {\bibinfo {author} {\bibfnamefont {G.}~\bibnamefont
  {Lazarides}}\ and\ \bibinfo {author} {\bibfnamefont {C.}~\bibnamefont
  {Panagiotakopoulos}},\ }\href {\doibase 10.1103/PhysRevD.52.R559} {\bibfield
  {journal} {\bibinfo  {journal} {Phys. Rev.}\ }\textbf {\bibinfo {volume}
  {D52}},\ \bibinfo {pages} {R559} (\bibinfo {year} {1995})},\ \Eprint
  {http://arxiv.org/abs/hep-ph/9506325} {arXiv:hep-ph/9506325} \BibitemShut
  {NoStop}%
\bibitem [{\citenamefont {Lazarides}\ \emph {et~al.}(1996)\citenamefont
  {Lazarides}, \citenamefont {Panagiotakopoulos},\ and\ \citenamefont
  {Vlachos}}]{Lazarides:1996rk}%
  \BibitemOpen
  \bibfield  {author} {\bibinfo {author} {\bibfnamefont {G.}~\bibnamefont
  {Lazarides}}, \bibinfo {author} {\bibfnamefont {C.}~\bibnamefont
  {Panagiotakopoulos}}, \ and\ \bibinfo {author} {\bibfnamefont {N.~D.}\
  \bibnamefont {Vlachos}},\ }\href {\doibase 10.1103/PhysRevD.54.1369}
  {\bibfield  {journal} {\bibinfo  {journal} {Phys. Rev.}\ }\textbf {\bibinfo
  {volume} {D54}},\ \bibinfo {pages} {1369} (\bibinfo {year} {1996})},\ \Eprint
  {http://arxiv.org/abs/hep-ph/9606297} {arXiv:hep-ph/9606297} \BibitemShut
  {NoStop}%
\bibitem [{\citenamefont {Jeannerot}\ \emph {et~al.}(2001)\citenamefont
  {Jeannerot}, \citenamefont {Khalil},\ and\ \citenamefont
  {Lazarides}}]{Jeannerot:2001qu}%
  \BibitemOpen
  \bibfield  {author} {\bibinfo {author} {\bibfnamefont {R.}~\bibnamefont
  {Jeannerot}}, \bibinfo {author} {\bibfnamefont {S.}~\bibnamefont {Khalil}}, \
  and\ \bibinfo {author} {\bibfnamefont {G.}~\bibnamefont {Lazarides}},\ }\href
  {\doibase 10.1016/S0370-2693(01)00429-4} {\bibfield  {journal} {\bibinfo
  {journal} {Phys. Lett.}\ }\textbf {\bibinfo {volume} {B506}},\ \bibinfo
  {pages} {344} (\bibinfo {year} {2001})},\ \Eprint
  {http://arxiv.org/abs/hep-ph/0103229} {arXiv:hep-ph/0103229} \BibitemShut
  {NoStop}%
\bibitem [{\citenamefont {Bastero-Gil}\ and\ \citenamefont
  {Berera}(2009)}]{BasteroGil:2009ec}%
  \BibitemOpen
  \bibfield  {author} {\bibinfo {author} {\bibfnamefont {M.}~\bibnamefont
  {Bastero-Gil}}\ and\ \bibinfo {author} {\bibfnamefont {A.}~\bibnamefont
  {Berera}},\ }\href {\doibase 10.1142/S0217751X09044206} {\bibfield  {journal}
  {\bibinfo  {journal} {Int.J.Mod.Phys.}\ }\textbf {\bibinfo {volume} {A24}},\
  \bibinfo {pages} {2207} (\bibinfo {year} {2009})},\ \Eprint
  {http://arxiv.org/abs/0902.0521} {arXiv:0902.0521 [hep-ph]} \BibitemShut
  {NoStop}%
\end{thebibliography}%

\end{document}